\documentstyle[sprocl,twoside,epsf,amsmath,amssymb,psfig,graphicx]{article} 

\newtheorem{definition}{Definition}[section]

\def\be{\begin{equation}}
\def\ee{\end{equation}}
\def\bea{\begin{eqnarray}}
\def\eea{\end{eqnarray}}

\def\slasha#1{\setbox0=\hbox{$#1$}#1\hskip-\wd0\hbox to\wd0{\hss\sl/\/\hss}}
\def\slashb#1{\setbox0=\hbox{$#1$}#1\hskip-\wd0\dimen0=5pt\advance
       \dimen0 by-\ht0\advance\dimen0 by\dp0\lower0.5\dimen0\hbox
         to\wd0{\hss\sl/\/\hss}}
\def\dd{{\mathrm{d}}}
\def\di{{\mathrm{i}}}
\def\de{{\mathrm{e}}}
\def\dM{{\mathbb{M}}}

\def\citebk#1{\hspace{0.9mm}\raisebox{-1.85mm}[0mm][0mm]
  {\Large\cite{#1}}\hspace{-0.1mm}}

\begin{document}
\sloppy

\title{SUPERMANIFOLDS -- APPLICATION TO SUPERSYMMETRY}

\author{PIERRE CARTIER}

\address{DMA, Ecole Normale Sup{\'e}rieure,\\
75005 Paris, France}

\author{C{\'E}CILE DeWITT-MORETTE, MATTHIAS IHL and CHRISTIAN S{\"A}MANN}

\address{University of Texas,\\ Department of Physics and Center for Relativity\\
Austin, TX 78712, USA} 


\maketitle

\begin{center}
Appendix by Maria E. Bell\\
\end{center}
\vspace{0.5cm}

\abstracts{Parity is ubiquitous, but not always
identified as a simplifying tool for computations.  
Using parity, having in mind the example of the bosonic/fermionic Fock
space, and the framework of ${\mathbb{Z}}_2$-graded (super) algebra, we
clarify relationships between the different definitions of supermanifolds 
proposed by various people. In addition, we work with four complexes allowing
an invariant definition of divergence:
\begin{itemize}
\item an ascending complex of forms, and a descending complex of
densities on real variables
\item an ascending complex of forms, and descending complex of
densities on Gra\ss mann variables.
\end{itemize}
This study is a step towards an invariant definition of integrals 
of superfunctions defined on supermanifolds leading to an 
extension to infinite dimensions. An application is given to a 
construction of supersymmetric Fock spaces.}

\vspace{0.4cm}


\tableofcontents

\newpage

\section{Dedication}
\setcounter{equation}{0}

\noindent {\textbf{Three quotes from M.S. Marinov.}} \\[0.3cm]
\noindent 1.  Particle spin dynamics as the Gra\ss mann variant of classical
mechanics, F.A.~Berezin and M.S.~Marinov, {\it Annals of Physics}
\textbf{104}, 336--362 (1977).
\begin{quote}
During the past few years a not so familiar concept has emerged in
high-energy physics, that of ``anticommuting
$c$-numbers.''\,\footnote{~Nowadays the expression ``$c$-numbers'' is
used for the even elements of the Gra\ss mann algebra, forming
a commutative algebra.}  The
formalism of the Gra\ss mann algebra is well known to mathematicians and
has been used for a long time.  The analysis on the Gra\ss mann algebra
was developed and exploited in a systematic way in applying the
generating functional method to the theory of second
quantization.\cite{11}\ldots
Seemingly,\footnote{~See quote \#3 from {\it
Physics Reports.}} the first physical work dealing with the
anticommuting numbers in connection with fermions was that by Matthews
and Salam.\cite{12}
\end{quote}
The authors mention also the works of Gervais and Sakita,\cite{13}
and Iwasaki and Kikkawa.\cite{14} \\[0.3cm]
2.  Classical spin and Gra\ss mann algebra, F.A.~Berezin and 
M.S.~Marinov, {\it JETP Lett.}, \textbf{21}, 320--321 (1975).
\begin{quote}
In view of the introduction of transformation groups with
anticommuting parameters into the theory of elementary
particles,\cite{15} and the intensive discussion of ``supersymmetry''
(see, for example, Zumino's review\,\cite{16}), universal interest has
been advanced in classical anticommuting quantities, i.e., in the
Gra\ss mann-algebra formalism. 
\end{quote}
3.  Path integrals in quantum theory, M.S. Marinov, {\it Physics
Reports}, \textbf{60}, 1--57 (1980).
\begin{quote}
As it was suggested by Schwinger\,\cite{17} (for further discussion see
the book\,\cite{18}), the matrix determining the Poisson brackets and the
canonical commutation relations, must be skew Hermitian; then the
consistent classical and quantum dynamics may be constructed on the
basis of the variation principle.  Remarkably, not only a real and
skew-symmetrical matrix (as in subsection 5.1), but also an imaginary
symmetrical one is possible.  However, in the second case the
canonical variables should be anticommuting.  Seemingly, this
suggestion did not attract attention in that time, though it is very
essential for understanding Schwinger's approach to quantum
electrodynamics.  The analysis in a space of anticommuting variables
(in the {\it Gra\ss mann algebra}) as exhaustively developed 
by,\cite{19}
and used for a consistent and unified functional approach to the
quantum theory with Fermi fields (see also the book by
Rzewusky\,\cite{19}).  The importance of using Gra\ss mann algebras was
essential for the discovery of supersymmetries and the recent
introduction of the superspace formalism.\\[0.2cm]
\noindent
\textbf{Acknowledgments} \\[0.2cm]
During my previous collaboration with Drs. F.~Berezin and M.~Terentyev,
I have benefited much, discussing with them some problems considered
in this report. \\
\textit{Note added in proof:} Having left ITEP forever with the
intention to settle in the Promised Land of my Fathers, I should like
to use this opportunity and to acknowledge gratefully the possibility
to work for many years at the Theoretical Divison, founded and
organised by the late Prof. Isaac Pomeranchuk.  I am much obliged to my
former colleagues, who have taught me many things; first of all, to
love the Science.
\end{quote}

\section{Preliminary Remarks}
\setcounter{equation}{0}

\subsection{What is Parity?}
Parity describes the behaviour of a product under exchange of its 
two factors. The so called Koszul's parity rule states:
\begin{center}
{\it ``Whenever you interchange two factors of parity 1, you 
get a minus sign.''}
\end{center}
Formally, this can be written as:
\begin{equation}
AB=(-1)^{\tilde{A}\tilde{B}}BA.
\end{equation}
where $\tilde{A}\in\{0,1\}$ denotes the parity of $A$.
\par We want this rule to be true for all kinds of commutative products, 
i.e.~products of Gra\ss mann variables, supernumbers, forms and
tensor densities, in particular
\begin{eqnarray}
A\wedge B&=&(-1)^{\tilde{A}\tilde{B}}B\wedge A.
\end{eqnarray}
Objects with parity 0 are called {\it even}, objects with parity 1 
{\it odd}.
\par For graded vectors and graded matrices, there exists no commutative 
product. The parity of a graded vector $X$ is given by its behaviour 
under multiplication with a graded scalar $z$:
\begin{equation}
z X=(-1)^{\tilde{z}\tilde{X}}X z
\end{equation}
A graded matrix is assigned even parity, if it preserves the parity of 
any graded vector under multiplication and odd, if it inverts the 
parity.
\par Graded vectors and graded matrices do not necessarily have a parity, 
but they can always be decomposed in a sum of a purely even and a 
purely odd part. 

\subsection{Why Gra\ss mann variables?}
The quotes from Marinov above aptly describe the birth of Gra\ss mann calculus 
in quantum field theory. We add a few comments. It is usually 
stated that the transition from a quantum mechanical system to a 
classical one is obtained by considering the limit 
$\hbar\rightarrow 0$. Hence the quantum mechanical relation
\begin{equation}
[q,p]=\di \hbar
\end{equation}
gives the commutativity $qp=pq$ in the limit.
\par In quantum field theory, the canonical quantization rules 
are\,\footnote{~As usual, $[A,B]$ is the commutator $AB-BA$ while 
$\{A,B\}$ is the anticommutator $AB+BA$.}
\begin{eqnarray}
\mbox{(bosonic case) }[\Phi(x),\Pi(y)]&=&\di \hbar\delta(x-y),\\
\mbox{(fermionic case) 
}\{\psi(x),\pi(y)\}&=&\di \hbar\delta(x-y).
\end{eqnarray}
In the (bosonic) case of the electromagnetic field, the 
Planck constant $\hbar$ enters in  the commutation rule, but not 
in the Lagrangian or the field equations. Hence the classical 
electromagnetism, with commuting field observables, obtains in the 
limit $\hbar\rightarrow 0$.
\par However, the (fermionic) case of Dirac electron 
field\,\footnote{~We drop spinor indices.} $\psi(x)$ is more subtle.
The normalization of the field is achieved by the 
requirement $j^\mu=e\bar{\psi}\gamma^\mu\psi$ for the electronic 
current (which solved the long-standing difficulties of the 
classical electron theory of Lorentz). Here we use the metric 
tensor $(\eta^{\mu\nu})={\mathrm{diag}}(-1,1,1,1)$ with
\begin{equation}
\gamma^\mu\gamma^\nu+\gamma^\nu\gamma^\mu=-2\eta^{\mu\nu}
\end{equation}
In the Dirac representation, the $\gamma^\mu$ are $4\times4$ 
matrices with $\gamma^0$ hermitian and $\gamma^1$, $\gamma^2$, 
$\gamma^3$ antihermitian. Finally 
$\bar{\psi}=\psi^\dagger\gamma^0$ is the charge conjugate spinor 
to $\psi$.
\par The Dirac equation is derived from the Lagrangian
\begin{equation}
{\mathcal{L}}=\bar{\psi}(-p_\mu\gamma^\mu-mc)\psi
\end{equation}
In the quantization, we use Schr\"odinger's Ansatz 
$p_a=-\di \hbar\partial_a$ for the 3-momentum, hence 
$p_\mu=-\di \hbar\partial_\mu$ for the relativistic 
4-momentum.\footnote{~Notice that $E/c=p^0=-p_0$ for the energy $E$, 
hence $E$ is quantized by $+\di \hbar\partial_t$ as it is 
done in the Schr\"odinger equation.} The Lagrangian becomes:
\begin{equation}
{\mathcal{L}}=\di \hbar\bar{\psi}\gamma^\mu\partial_\mu\psi-mc\bar{\psi}\psi
\end{equation}
hence the canonical conjugate momentum
\begin{equation}
\pi(x)=\frac{\delta\mathcal{L}}{\delta\dot{\psi}}=\di \hbar\psi^*(x)
\end{equation}
The canonical anticommutation relations are now
\begin{equation}
\{\psi(x),\psi^*(y)\}=\delta(x-y)
\end{equation}
and remain identical in the limit $\hbar\rightarrow 0$, and don't 
give rise to Gra\ss mann quantities\,\footnote{~ This result is 
not a fermionic artefact, as for example the full Lagrangian for the 
Klein-Gordon field gives rise to the commutation relation 
$[\Phi(x),\dot\Phi(y)]=\frac{1}{\hbar}\delta(x-y)$, which has no 
classical limit for $\hbar\rightarrow 0$ as well.}. 
\par The reason for using Gra\ss mann variables is not so much the 
development of a pseudo-classical mechanics but rather the need of 
representing the algebra of fermionic position and momentum 
observables as operators on a space of functions. This is easily 
achieved using functions of Gra\ss mann variables: 
\begin{equation}
\{\xi,\xi\}=0,\hspace{1cm}
\{-\di \hbar\frac{\partial}{\partial\xi},\di \hbar\frac{\partial}{\partial\xi}\}=0,\hspace{1cm}
\{\xi,i\hbar\frac{\partial}{\partial\xi}\}=i\hbar
\end{equation}
Note that in the bosonic case, the commutation relation 
$[q,q]=qq-qq=0$ does not contain any information. The only 
nontrivial rule is $[q,p]=\di \hbar$ contrary to the 
fermionic case, where already the rule $\{q,q\}=qq+qq=0$ is 
nontrivial and demands the use of anticommuting objects.
\par To us, the most convincing reason for using Gra\ss mann variables is 
that we need them to use the path integral method in the case of 
fermionic fields. This is clearly stated in the quotations above.

\newpage
\part{Foundations}
\pagestyle{myheadings}  
\markboth{\quad\small \em P. Cartier, C. DeWitt-Morette, M. Ihl and C. S{\"a}mann \hfill}{\hfill \small \em Supermanifolds -- application to supersymmetry\quad}
\section{Definitions and notations}
\setcounter{equation}{0}

\begin{itemize}
\item Basic graded algebra ~~~ $\tilde{A} := \mbox{parity of }A \in
\{0\,, 1\}$ \\
Parity of a product: 
$\widetilde{AB}=\tilde{A}+\tilde{B}{\mbox{ mod }} 2$.\\
Graded commutator\,\footnote{~Another notation is $[A,B]_\mp=AB\mp BA$.}
$[A\,, B] := AB - (-1)^{\tilde{A}\tilde{B}}BA$\\
Graded anticommutator $\{A\,, B\} := AB +
(-1)^{\tilde{A}\tilde{B}}BA$\\
Graded Leibnitz rule 
$$ D(A\cdot B) = DA \cdot B + (-1)^{\tilde{A}\tilde{D}}A \cdot DB$$
previously called ``antiLeibnitz'' when $\tilde{D} = 1$ \\
Graded symmetry $A^{\cdots\alpha\beta\cdots} =
(-1)^{\tilde{\alpha}\tilde{\beta}} A^{\cdots\beta\alpha\cdots}$ \\
Graded antisymmetry $A^{\cdots\alpha\beta\cdots} =
-(-1)^{\tilde{\alpha}\tilde{\beta}}A^{\cdots\beta\alpha\cdots}$ \\
Graded Lie derivative ${\mathcal{L}}_{X} = [\di_{X}\,, \dd]_{+}\,,
{\mathcal{L}}_{\Xi} = [\di_{\Xi}\,, \dd]_{-}$

\item Supernumbers \\
Gra\ss mann generators $\{\xi^{\mu}\} \in \Lambda_{\nu}\,,
\Lambda_{\infty}\,, \Lambda$
\\$\xi^{\mu}\xi^{\sigma} = -\xi^{\sigma}\xi^{\mu}\,;$ $\Lambda =
\Lambda^{\mathrm{even}} \oplus  \Lambda^{\mathrm{odd}}$ \\
Supernumber $z = u + v$, $u$ is even, $v$ is odd \\
\hspace*{.4in} $z = z_{B} + z_{S}$, ~$z_{B} \in {\mathbb{R}}$
is the body, $z_{S}$ is the soul\\
Complex conjugation of supernumber: $(zz^{\prime})^{\ast} = z^{\ast}z^{\prime\ast}$ 
(see appendix).

\item Superpoints\\
Real coordinates $x\,,y \in {\mathbb{R}}^{n}$,~~$x = (x^{1}\,, \cdots\,, x^{n})$ \\
$\left( x^{1}\,, \ldots\,, x^{n}\,, \xi^{1}\,, \ldots, \xi^{\nu}\right)
\in {\mathbb{R}}^{n|\nu}$, condensed notation $x^{A} = (x^{a}\,, \xi^{\alpha})$
\\
$\left( u^{1}\,, \ldots\,, u^{n}\,, v^{1}\,,\ldots\,, v^{\nu} \right)
\in {\mathbb{R}}^{n}_{c} \times {\mathbb{R}}^{\nu}_{a}$ 

\item Supervectorspace: (graded) module over the ring of supernumbers \\
$X = U + V$, $U$ even, $V$ odd \\
$X = e_{(A)}{}^{A}X$ \\
$X^A=(-1)^{\tilde{X}\tilde{A}}\,{}^AX$\\
The even elements of the basis $(e_{(A)})_A$ are listed first. A
supervector is even if each of its coordinates ${}^{A}X$ has the same
parity as the corresponding basis element $e_{(A)}$. It is odd if the
parity of ${}^{A}X$ is opposite to the parity of $e_{(A)}$.  Parity
cannot be assigned in other cases.

\item Graded Matrices \\
Four different uses of graded matrices\\
\begin{tabular}{l c}
given $V = e_{(A)} {}^{A}V = \bar{e}_{(B)} {}^{B} \bar{V}$ with $A = (a\,,
\alpha)$ and & $e_{(A)} =
\bar{e}_{(B)} {}^{B}M_{A}$\\[0.1cm]
\hspace*{2.7in} then & ${}^{B}\bar{V} = {}^{B}M_{A} {}^{A}V$ \\[0.1cm]
\end{tabular}
given $\langle \omega\,, V\rangle = \omega_{A} {}^{A}V = \bar{\omega}_{B}
{}^{B}\bar{V}$ where $\omega = \omega_{A} {}^{(A)}\theta =
\bar{\omega}_{B} {}^{(B)}\bar{\theta}$ \\[0.1cm]
then $\langle \omega\,, V\rangle = \omega_{A} \langle
{}^{(A)}\theta\,, e_{(B)} \rangle {}^{B}V$ implies $\langle
{}^{(A)}\theta\,, e_{(B)} \rangle = {}^A\delta_{B}$, \\[0.1cm]
\begin{tabular}{l c}
 & $\omega_{A} = \bar{\omega}_{B} {}^{B} M_{A}$, \\[0.1cm]
\hspace*{2.7in} and & ${}^{(B)}\bar{\theta} = {}^{B}M_{A}{}^{(A)} \theta$
\end{tabular}\\[0.1cm]
Matrix parity \\
$\tilde{M} = 0$, if $\forall A ~\mbox{and}~ B, ~
\widetilde{{}^{B}M_{A}} + \widetilde{\mbox{column}B} +
\widetilde{\mbox{row} A} = 0 \bmod 2$. \\
$\tilde{M} = 1$, if $\forall A~\mbox{and}~ B, ~
\widetilde{{}^{B}M_{A}} + \widetilde{\mbox{column}B} +
\widetilde{\mbox{row}A} = 1\bmod 2$.
\\By multiplication, an even matrix preserves the parity of the vector components, an 
odd matrix inverts the parity of the vector components.

\item Parity assignments \\
$\tilde{\dd} = 1$ ~ $(\widetilde{\dd x}) = \tilde{\dd} + \tilde{x} = 1$ ~~~
$(\widetilde{\dd\xi}) = \tilde{d} + \tilde{\xi} = 0$ \\
($x$ ordinary variable, $\xi$ Gra\ss mann variable)\\
~~~~~~~ $(\widetilde{\partial/\partial x}) = \tilde{x} = 0$ ~~~
$(\widetilde{\partial/\partial\xi}) = \tilde{\xi} = 1$ \\
$\tilde{\di} = 1$ ~ $\widetilde{\di_{X}} = \tilde{\di} + \tilde{X} = 1$ ~
$\widetilde{\di}_{\Xi} = \tilde{\di} + \tilde{\Xi} = 0$ \\
Parity of real $p$-forms: even for $p=0 \bmod 2$, odd for $p=1 \bmod 2$ \\
Parity of Gra\ss mann $p$-forms: always even.\\
Graded exterior product $\omega \wedge \eta =
(-1)^{\tilde{\omega}\tilde{\eta}}\eta \wedge \omega$
\end{itemize}

\subsection{Supernumbers}
We shall work with a Gra\ss mann algebra with generators 
$\xi^\mu$; we can assume that their collection is finite 
$\xi^1,...,\xi^N$ for some integer $N\geq 1$, or infinite 
$\xi^1,\xi^2,...$ The Gra\ss mann algebra is denoted by 
$\Lambda_N$ in the first case and $\Lambda_\infty$ in the second. 
If we want to remain ambiguous, we use the notation $\Lambda$ to 
mean either $\Lambda_N$ or $\Lambda_\infty$.
\par The generators $\xi^\mu$ anticommute 
$\xi^\mu\xi^\nu=-\xi^\nu\xi^\mu$ and in particular 
$(\xi^\mu)^2=0$. Hence an arbitrary supernumber $z$ can be written 
uniquely in the form $\sum_{p\geq 0}z_p$ where each $z_p$ is of 
the form
\begin{equation}
z_p=\frac{1}{p!}z_{\alpha_1...\alpha_p}\xi^{\alpha_1}...\xi^{\alpha_p};
\end{equation}
the coefficients $z_{\alpha_1...\alpha_p}$ are antisymmetrical in 
the indices $\alpha_1,...,\alpha_p$ and may be real or complex 
(see appendix).
\par We can split $z$ as
\begin{eqnarray}
z=u+v&{\mathrm{with}}&u=z_0+z_2+...\nonumber\\
&&v=z_1+z_3+...\nonumber\\
z=z_B+z_S&{\mathrm{with}}&z_B=z_0\nonumber\\
&&z_S=z_1+z_2+...\nonumber
\end{eqnarray}
Hence $u$($v$) is called even (odd) since it involves products of 
even (odd) number of generators. Furthermore, the real number 
$z_B$ is called the body and $z_S$ the soul of $z$; in the case of 
$\Lambda_N$, $z_S$ is nilpotent, i.e. $(z_S)^{N+1}=0$.

\subsection{Superspaces}
Throughout our work with supermanifolds we encounter three 
different graded spaces:
\begin{equation}
{\mathbb{R}}^{n|\nu}\hspace{1cm}\Lambda^{n+\nu}\hspace{1cm}{\mathbb{R}}_c^n\times{\mathbb{R}}_a^\nu\nonumber
\end{equation}
All these spaces have $n+\nu$ coordinates, but of different origin:
\begin{eqnarray}
{\mathbb{R}}^{n|\nu}&:& (x^1,...,x^n,\xi^1,...,\xi^\nu)\hspace{0.3cm}x^i\in{\mathbb{R}},\hspace{0.3cm} \xi^i \mbox{ Gra\ss mann variables}\nonumber\\ 
\Lambda^{n+\nu}&:& (z^1,...,z^{n+\nu})\hspace{0.3cm} z^i\in\Lambda\nonumber\\
{\mathbb{R}}_c^n\times{\mathbb{R}}_a^\nu&:& (u^1,...,u^n,v^1,...,v^\nu)\hspace{0.3cm}u^i\in\Lambda^{\mathrm{even}},\hspace{0.3cm}v^i\in\Lambda^{\mathrm{odd}}\nonumber
\end{eqnarray}
We want to compare these spaces to vectorspaces and thus associate a basis to each one:
\begin{eqnarray}
{\mathbb{R}}^{n|\nu}&:& \mbox{ $n+\nu$ {\it non}-graded elements: }(e_1,...,e_{n+\nu})\nonumber\\
\Lambda^{n+\nu}&:& \mbox{ pure basis: $n$ even and $\nu$ odd elements: }(e_1,...,e_n,\epsilon_1,...,\epsilon_\nu)\nonumber\\
{\mathbb{R}}_c^n\times{\mathbb{R}}_a^\nu&:& \mbox{ pure basis: $n$ even and $\nu$ odd elements: }(e_1,...,e_n,\epsilon_1,...,\epsilon_\nu)\nonumber
\end{eqnarray}
The first space is obviously isomorphic to a real graded 
vectorspace. The second one is a supervectorspace as defined 
e.g.~by B.S. DeWitt. The last space is the subset of the {\it even} 
elements of the supervectorspace. Thus it is {\it not} a 
supervectorspace, since this set is not closed under multiplication 
with an odd supernumber.

\subsection{Superfunctions}\label{superfunctions}

A superfunction is a function depending on even and 
odd variables. Since the square of an odd
object vanishes, the Taylor expansion in the odd variables has 
only finitely many terms:
\begin{eqnarray}
f(x^1,...,x^n,\xi^1,...,\xi^\nu)&=&f_0+f_1\xi^1+...
+f_{12}\xi^1\xi^2
+...+f_{1...\nu}\xi^1...\xi^\nu\nonumber
\end{eqnarray}
where the coefficients are functions of the even variables: 
$f_I=f_I(x^1,...,x^n)$.
\par We will now have a closer look at superfunctions 
$f:\Lambda\rightarrow\Lambda$. In analogy to the complex case, we 
call a superfunction analytic, if it has an expansion as follows:
\begin{equation}\label{superanalytic}
f(z)=\sum_n \alpha_n z^n
\end{equation}
where $\alpha_n\in\Lambda$.
Given an analytic complex function 
$f_{\mathbb{C}}:{\mathbb{C}}\rightarrow{\mathbb{C}}$, we want to 
examine how to continue it to a superanalytic function on 
$\Lambda$. Having in mind the expression for the inverse of a 
supernumber
\begin{equation}
z^{-1}=z_B^{-1} \sum_{n}\left( -\frac{z_S}{z_B} \right) ^n=
\sum_{n} \frac{(-1)^n}{z_B^{n+1}} z_S^n
\end{equation}
we rewrite the expansion (\ref{superanalytic}) to a similar form:
\begin{equation}
f(z)=\sum_n \alpha_n z^n=\sum_n \alpha_n (z_B+z_S)^n=\sum_n \alpha'_n 
z_S^n.
\end{equation}
Comparing the coefficients, we note that the formula for the 
continuation of a complex analytic function to a superanalytic 
function should be given by
\begin{equation}\label{analyticcontinued}
f(z)=\sum_{n=0}^\infty \frac{1}{n!}f_{\mathbb{C}}^{(n)}(z_B)z_S^n.
\end{equation}
where $f_{\mathbb{C}}^{(n)}$ is the $n$-th derivative of 
$f_{\mathbb{C}}$. A general superanalytic function then obviously has the form
\begin{equation}\label{genanfunc}
f(z)=\sum_{n=0}^\infty 
f_{\alpha_1...\alpha_n}(z)\xi^{\alpha_1}...\xi^{\alpha_n}
\end{equation}
where the $f_{\alpha_1...\alpha_n}(z)$ are functions of the form 
(\ref{analyticcontinued}) and the $\xi^i$ are Gra\ss mann generators.
\par With some trivial algebra, we can consider $f(z)$ as a function 
of an even and an odd supernumber: $f(z)=f(u,v)$ and rewrite 
(\ref{analyticcontinued}):
\begin{equation}\label{upvelts}
f(u,v)=f(u)+g(u)v=\left(\sum_{n=0}^\infty 
\frac{1}{n!}f_{\mathbb{C}}^{(n)}(u_B)u_S^n\right)+\left(\sum_{n=0}^\infty 
\frac{1}{n!}g_{\mathbb{C}}^{(n)}(u_B)u_S^n\right)v
\end{equation}
where $f_{\mathbb{C}}$ and $g_{\mathbb{C}}$ are complex 
functions. A general superanalytic function is again
\begin{equation}\label{genanupvfunc}
f(u+v)=\sum_{n=0}^\infty 
f_{\alpha_1...\alpha_n}(u+v)\xi^{\alpha_1}...\xi^{\alpha_n}
\end{equation}
now with $f_{\alpha_1...\alpha_n}(u+v)$ functions of the form 
(\ref{upvelts}). 
\par These concepts can easily be generalized to functions 
$f:{\mathbb{C}}^n_c\times{\mathbb{C}}^\nu_a\rightarrow \Lambda$ 
which take the form:
\begin{equation}
f(u^1,...,u^n,v^1,...,v^\nu) = \sum_{r=0}^\nu\sum_{s=0}^\infty c
\frac{\partial^sF^m_{\mu_1...\mu_r}(u_B)}{\partial
u_B^{n_1}...\partial u_B^{n_s}}u_S^{n_1}...x_S^{n_s}v^{\mu_r}...v^{\mu_1}
\end{equation}
with a complex constant $c$. For $F^m_{\mu_1...\mu_r}$, we allow 
functions ${\mathbb{C}}\rightarrow\Lambda$ instead of ${\mathbb{C}}\rightarrow{\mathbb{C}}$
which implies the generalizing step, for example, from (\ref{upvelts}) to 
(\ref{genanupvfunc}).
\par If we now put constrains on the $F^m_{\mu_1...\mu_r}$, we can 
construct arbitrary superanalytic functions $f:{\mathbb{R}}^n_c\times{\mathbb{R}}^\nu_a\rightarrow \Lambda^{\mathrm{even}}$
or $f:{\mathbb{R}}^n_c\times{\mathbb{R}}^\nu_a\rightarrow 
\Lambda^{\mathrm{odd}}$ which we will use for coordinate 
transformations on supermanifolds.

\section{Supermanifolds and Sliced Manifolds}
\setcounter{equation}{0}

\subsection{Definition}
Ordinary manifolds are topological spaces which are locally 
diffeomorphic to ${\mathbb{R}}^n$. To generalize manifolds to B.S.
DeWitt's construction of supermathematics, it seems obvious to 
choose $\Lambda^n$ for replacing ${\mathbb{R}}^n$, but this choice has 
several shortcomings: we lose the notion of even and odd 
components of the supermanifold and thus the ${\mathbb{Z}}_2$-grading. 
Furthermore, even if we split $\Lambda^n$ in 
$(\Lambda^{\mathrm{even}})^n\times(\Lambda^{\mathrm{odd}})^n$, the 
number of even and odd dimensions is always equal. 
\par Contrary to $\Lambda^n$, ${\mathbb{R}}_c^n\times{\mathbb{R}}_a^\nu$ 
has none of the disadvantages above. Before we can use 
this space, we first have to introduce a topology. We can use the one 
induced from ordinary ${\mathbb{R}}^n$:
\vspace{0.2cm}
\begin{definition}
Let $\pi:{\mathbb{R}}_c^n\times{\mathbb{R}}_a^\nu\rightarrow{\mathbb{R}}^n$ be the projection
\begin{equation}
\pi(x^1,...,x^n,\eta^1,...,\eta^\nu):=(x_B^1,...,x_B^n).
\end{equation}
A set $X\subset{\mathbb{R}}_c^n\times{\mathbb{R}}_a^\nu$ is called open, 
if there is a set $Y\subset{\mathbb{R}}$ with $X=\pi^{-1}(Y)$.
\end{definition}
\vspace{0.2cm}
Obviously, this space is no longer hausdorff, but 
only projectively hausdorff\,\footnote{~Two points can only be 
contained in two open set with empty intersection and each in one, if their 
coordinate tuples have different bodies.}.
Furthermore, one should keep in mind 
that ${\mathbb{R}}_c^n\times{\mathbb{R}}_a^\nu$ is not a supervector 
space as it is not closed under multiplication with an odd 
supernumber.
\par Now the definition of a supermanifold is straightforward:
\vspace{0.2cm}
\begin{definition}
A {\bf supermanifold} $\dM$ is a topological space which is locally 
diffeomorphic to ${\mathbb{R}}_c^n\times{\mathbb{R}}_a^\nu$. The 
dimension of $\dM$ will be denoted by $(n,\nu)$. 
\end{definition}

\subsection{Body, Aura and Equivalence Classes}
We defined the body of a supernumber as its purely real or complex component. 
A similar structure can be introduced on supermanifolds. Such a structure has 
to be invariant under the general coordinate transformation (CT1):
\begin{eqnarray}
\bar{x}^m & = & \sum_{r=0}^\nu\sum_{s=0}^\infty c_1
\frac{\partial^sX^m_{\mu_1...\mu_r}(x_B)}{\partial
x_B^{n_1}...\partial x_B^{n_s}}x_S^{n_1}...x_S^{n_s}\eta^{\mu_r}...\eta^{\mu_1}\\
\bar{\eta}^\mu & = & \sum_{r=0}^\nu\sum_{s=0}^\infty c_2
\frac{\partial^sX^\mu_{\mu_1...\mu_r}(x_B)}{\partial
x_B^{n_1}...\partial x_B^{n_s}}x_S^{n_1}...x_S^{n_s}\eta^{\mu_r}...\eta^{\mu_1}
\end{eqnarray}
as developed in section \ref{superfunctions}, 
$c_1,c_2\in{\mathbb{C}}$.
This forbids the following naive definition of a body:
\\ ``Let $\dM$ a supermanifold $\dM$ with a chart
$\phi$ mapping $\dM$ on ${\mathbb{R}}^n_c\times{\mathbb{R}}^\nu_a$ and the
map $b(x^1,...,x^n,\eta^1,...,\eta^\nu):=(x_B^1,...,x_B^n,0,...,0)$. 
Then for a point $p$ in $\dM$ its body is given by $\phi^{-1}\circ
b\circ\phi(p)$."
\begin{figure}[h]
\centerline{\includegraphics[angle=-90,width=12.5cm,totalheight=8cm] {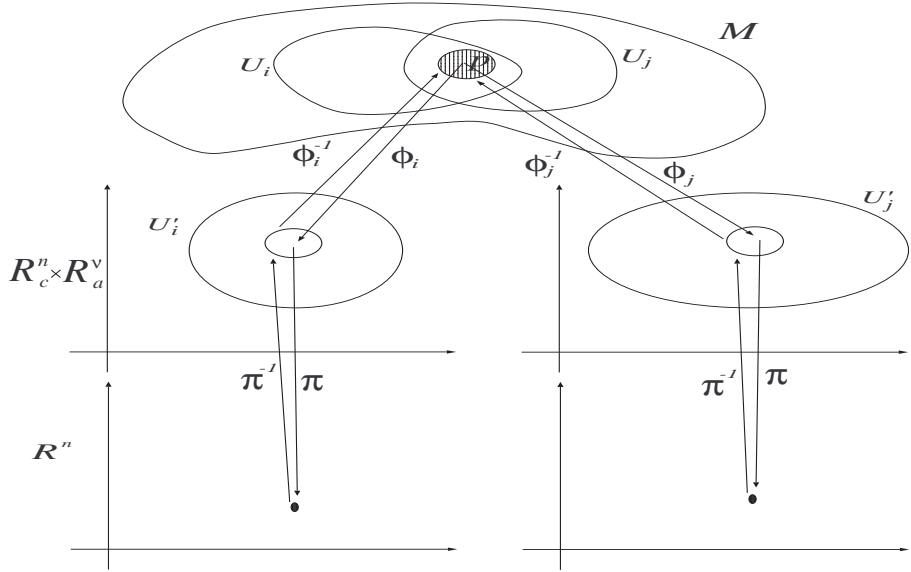}}
\caption[]{Aur\ae{}, as the one shown here (shaded area), are invariant under coordinate transformation:
$\phi^{-1}_i\circ\pi^{-1}\circ\pi\circ\phi_i=\phi^{-1}_j\circ\pi^{-1}\circ\pi\circ\phi_j$. This
enables us to define the body of a supermanifold.}
\end{figure}
\par Instead, we have to introduce the term ``aura of a point'':
\vspace{0.2cm}
\begin{definition}
Given a supermanifold $\dM$ with chart $\phi$ mapping $\dM$ on
${\mathbb{R}}^n_c\times{\mathbb{R}}^\nu_a$, then the set
$A(x):=\phi^{-1}\circ\pi^{-1}\circ\pi\circ\phi(x)$ for any $x\in \dM$ 
is called the {\bf aura}\,\footnote{~This set is also called ``soul subspace".} of x.
\end{definition}
\vspace{0.2cm}
The aura of a point is invariant under coordinate transformation, 
i.e.~$\bar{x}(A(x))=A(\bar{x}(x))$ which is short for 
\begin{equation}
\forall p\in A(x):\bar{x}(p)\in A(\bar{x}(x)).
\end{equation}
Considering aur\ae{} as points of a real manifold, we find the 
invariant definition for the body of a supermanifold:
\vspace{0.2cm}
\begin{definition}
The real manifold $\dM_B=\{A(x)|x\in \dM\}$ with chart $\pi\circ\phi$ mapping $\dM$ on
${\mathbb{R}}^n$ is called the {\bf body} of $\dM$.
\end{definition}
\vspace{0.2cm}
In this definition, points of a supermanifold which belong to the 
same aura are not distinguished. So it is natural to introduce an 
equivalence class of points:
\vspace{0.2cm}
\begin{definition}
Two points on a supermanifold are called equivalent, if and only
if they have the same aura: $x\sim y\Leftrightarrow A(x)=A(y)$.
\end{definition}

\subsection{Sliced Manifolds}
Given a supermanifold $\dM$ together with the equivalence relation 
above, we can choose a representative for each equivalence class.
Together with the chart $\phi_S=\pi\circ\phi$, the set of 
representatives can be considered as a real manifold $M'$. Attaching 
at each point $p\in \dM'$ all equivalent points $\{p'\sim p|p\in \dM\}$ 
as a fiber leads to the picture of a sliced manifold $\dM_o$.
\begin{figure}[h]
\centerline{\includegraphics[angle=-90,width=11cm,totalheight=5cm] {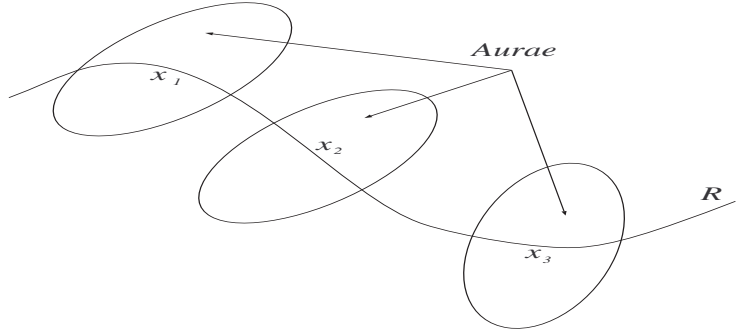}}
\caption[]{A sliced supermanifold consists of a real manifold $M'$, which is the set of chosen representatives
for the aur\ae{}. The fiber at a point $x$ is the set of equivalent points to the representative
$x$: $F_x=\{p|p\sim x\}$ and can thus be regarded as the aur\ae{} attached to each representative.}
\end{figure}
Note that given a sliced manifold $\dM_o$ constructed from $\dM$, each 
other sliced manifold $\dM_o'$ also constructed from $\dM$ is just a section 
of $\dM_o$, as it corresponds to a different choice of representatives 
of the equivalence classes or of elements of the aur\ae.
\par A sliced supermanifold is locally described by $n$ real
coordinates referring to the body of the supermanifold and $n$
even and $\nu$ odd coordinates, containing Gra\ss mann
generators: $(x^1,...,x^n,y^1,...,y^n,\eta^1,...,\eta^\nu)$. The
intrinsic coordinate transformation for such a sliced
supermanifold is (CT2):
\begin{eqnarray}
\bar{x}^m&=&X^m(x)\\
\bar{y}^m & = & \sum_{r=0}^\nu\sum_{s=0}^\infty c_1
\frac{\partial^s Y^m_{\mu_1...\mu_r}(x)}{\partial
x^{n_1}...\partial x^{n_s}}y^{n_1}...y_S^{n_s}\eta^{\mu_r}...\eta^{\mu_1}\\
\bar{\eta}^\mu & = & \sum_{r=0}^\nu\sum_{s=0}^\infty c_2
\frac{\partial^s Y^\mu_{\mu_1...\mu_r}(x)}{\partial
x^{n_1}...\partial
x^{n_s}}y^{n_1}...y^{n_s}\eta^{\mu_r}...\eta^{\mu_1}.
\end{eqnarray}
The constants are the same as in (CT1), $X^m(x)$ is an arbitrary
bijective function, mapping real numbers to real numbers and
$Y^m$, $Y^\mu$ are functions\,\footnote{~Excluding ${\mathbb{R}}$ makes sure, that the
range consists only of supernumbers without body.}
${\mathbb{R}}^n\rightarrow{\mathbb{R}}_c\backslash{\mathbb{R}}$ or
${\mathbb{R}}^n\rightarrow{\mathbb{R}}_a$ so that
the parity of the equations is matched.
\par In a further step, we can linearize the slices. It is clear,
that varying $(y^i)_i$ and $(\eta^\iota)_\iota$ for one $x$ by bodiless
values, we obtain the whole slice at $x$. Since the coordinates are
bodiless themselves, multiplication with a supernumber does not change this
and we can consider this space as a supervector space over the ring of supernumbers.
The intrinsic coordinate transformations here (CT3) are linear maps:
\begin{eqnarray}
\bar{x}^m&=&X^m(x)\\
\bar{y}^m & = & Y(x)^m_ny^n+Y(x)^m_\nu \eta^\nu\\
\bar{\eta}^\mu & = & \Upsilon(x)^\mu_n y^n+\Upsilon(x)^\mu_\nu
\eta^\nu,
\end{eqnarray}
where $Y(x)^m_n$ and $\Upsilon(x)^\mu_\nu$ are maps
${\mathbb{R}}^n\rightarrow{\mathbb{R}}_c$ and $Y(x)^m_\nu$ and
$\Upsilon(x)^\mu_n$ are maps ${\mathbb{R}}^n\rightarrow{\mathbb{R}}_a$.

\subsection{Some sliced manifolds are Kostant manifolds}
We use the definition of a Kostant manifold given in Y. Choquet-Bruhat and 
C. DeWitt-Morette,\cite{Choquet-Bruhat}
\\A Kostant bundle $K$ over $\dM$ is a fiber bundle $K$ over a manifold $\dM$
where the bundle coordinates take their values in a graded algebra $A$ and the 
transition map from one chart to another commutes with the product of the algebra.
Now we can define:
\vspace{0.2cm}
\begin{definition}
A Kostant manifold is a pair $(M,K)$ where $\dM$ is an 
ordinary $C^\infty$ manifold and $K$ a Kostant bundle over M. A 
graded function is a section of the bundle.
\end{definition}
\vspace{0.2cm}
Let us consider a sliced manifold constructed as above from a 
supermanifold with dimension $(n,n)$. Locally, we can describe 
this object by $n$ real variables $x^i$, $n$ bodiless even 
variables $y^i$ taking their values in $\Lambda^{\mathrm{even}}$ and $n$ odd 
variables $\eta^i$ with values in $\Lambda^{\mathrm{odd}}$. Since the 
decomposition of a supernumber in its even and odd parts is unique, we 
can combine the bodiless coordinates, i.e.~the coordinates of the 
aura, to a new coordinate $z^i=y^i+\eta^i$ without losing any 
information. 
\par We obtain a real manifold with a bundle, which is a Gra\ss mann 
algebra without bodiless elements, i.e.~generated by the Gra\ss 
mann generators $\xi^i$, but the polynomial of degree zero is not 
included. 
\par This description of a sliced manifold is obviously also a Kostant 
manifold. In this sense, B.S. DeWitt's supermanifolds with equally many odd
and even dimensions can always be reduced to Kostant manifolds.

\section{Graded manifolds}
\setcounter{equation}{0}

\subsection{Basic definitions}
Graded manifolds are the most trivial ones discussed in literature.
Their definition is the obvious generalization of real manifolds, 
using ${\mathbb{R}}^{n|\nu}$ instead of ${\mathbb{R}}$. We 
follow basically the definition given by Voronov\,\cite{Voronov}.
\par The space ${\mathbb{R}}^{n|\nu}$ can be defined by the functions on this space
which take their values in the Gra\ss mann algebra $\Lambda_\nu$:
\begin{equation}
C^\infty({\mathbb{R}}^{n|\nu})=\{f_0+f_1\xi^1+...+f_{12}\xi^1\xi^2+...+f_{1...\nu}\xi^1...\xi^\nu
|f_I \in C^\infty({\mathbb{R}}^n)\}
\end{equation}
It is obvious that ${\mathbb{R}}^{n|\nu}$ can be described by 
coordinates 
\begin{equation}
x^A=(x^a,\xi^\alpha)=(x^1,...,x^n,\xi^1,...,\xi^\nu)
\end{equation}
where the $x^a$ and the $\xi^\alpha$ are real and Gra\ss mann 
variables respectively A possible topology of this space is induced by 
the real component as in the case of supermanifolds. Given the 
projection $\pi$ by
\begin{equation}
\pi(x^1,...,x^n,\xi^1,...,\xi^\nu):=(x^1,...,x^n),
\end{equation}
a subset $X\subset{\mathbb{R}}^{n|\nu}$ is called open, if and only 
if $X=\pi^{-1}(Y)$ where $Y$ is an open set in ${\mathbb{R}}^n$. As in the 
case of supermanifolds, with this topology ${\mathbb{R}}^{n|\nu}$ is only projectively 
hausdorff.
\par Now we are ready to define:
\vspace{0.2cm}
\begin{definition}
A graded manifold is a topological space which is locally diffeomorphic to
${\mathbb{R}}^{n|\nu}$. 
\end{definition}
\vspace{0.2cm}
Simple examples for graded manifolds are $\Pi T\dM$ and $\Pi T^*\dM$, the 
tangent and cotangent bundle with changed parity of the bundle 
coordinates.
\par In the case of graded manifolds, the definition of the body is much 
easier than for supermanifolds. Here we can use the projection
\begin{equation}
b(x^1,...,x^n,\xi^1,...,\xi^\nu)=(x^1,...,x^n,0,...,0)
\end{equation}
and the "naive" definition:
\vspace{0.2cm}
\begin{definition}
Given a graded manifold $\dM$ then its body $\dM_B$ is given by
\begin{equation}
\dM_B=\{b(x) | x\in \dM\}
\end{equation}
and its soul is $\dM_S=\dM\backslash \dM_B$.
\end{definition}
\vspace{0.1cm}
Since $(\xi^i)^2=0$, we can look at the soul $\dM_S$ of a graded manifold $\dM$ as 
infinitely many copies of $\dM_B$ infinitely close to $\dM_B$.

\vspace*{-0.2cm}
\subsection{From supermanifolds to graded manifolds via 
superfunctions}
Though supermanifolds seem to be much richer in their properties than 
graded manifolds, we are basically interested in the space of functions on
both. 
\par On supermanifolds, an arbitrary function can be expanded in 
polynomials of the odd variables:
\begin{equation}
f(Y,\Upsilon)=f_0(Y)+f_1(Y)\Upsilon^1+...+f_{12}(Y)\Upsilon^1\Upsilon^2+...+f_{1...\nu}(Y)\Upsilon^1...\Upsilon^\nu.
\end{equation}
Here the $f_I$ are functions 
${\mathbb{R}}_c^n\rightarrow\Lambda_\infty$.
\par The functions on graded manifolds are, as discussed above:
\begin{equation}
f(x,\xi)=f_0(x)+f_1(x)\xi^1+...+f_{12}(x)\xi^1\xi^2+...+f_{1...\nu}(x)\xi^1...\xi^\nu,
\end{equation}
where the $f_I$ are functions ${\mathbb{R}}^n\rightarrow{\mathbb{R}}$.
\par With these expansions, it is obvious that functions on 
supermanifolds and those on graded manifolds differ only in the 
possible values of the $f_I$. As real --- or complex --- degrees of 
freedom are sufficient for our purposes, we can constrain our 
considerations to graded manifolds without fearing to lose 
properties from supermanifolds.

\part{Integration}
\pagestyle{myheadings}  
\markboth{\quad\small \em P. Cartier, C. DeWitt-Morette, M. Ihl and C. S{\"a}mann \hfill}{\hfill \small \em Supermanifolds -- application to supersymmetry\quad}
\section{Definitions and notations}
\setcounter{equation}{0}

\begin{itemize}
\item Forms and densities of weight one on ${\mathbb{M}}^{D}$
(without metric tensor) \\
$({\mathcal{A}}^{\bullet}\,, \dd)$ Ascending complex of forms $\dd:
{\mathcal{A}}^{p} \rightarrow {\mathcal{A}}^{p+1}$ \\
$({\mathcal{D}}_{\bullet}\,, \nabla \cdot \mbox{ or } {\mathrm{b}})$ Descending
complex of densities $\nabla \cdot : {\mathcal{D}}_{p} \rightarrow
{\mathcal{D}}_{p-1}$ \\
${\mathcal{D}}_{p} \equiv {\mathcal{D}}^{-p}$ used for ascending
complex in negative degrees

\item Operators on ${\mathcal{A}}^{\bullet}({\mathbb{M}}^{D})$ \\
$M(f): {\mathcal{A}}^{p} \rightarrow {\mathcal{A}}^{p}$,
multiplication by a scalar function $f : {\mathbb{M}}^{D} \rightarrow
{\mathbb{R}}$ \\
$\de (f): {\mathcal{A}}^{p} \rightarrow {\mathcal{A}}^{p+1}$ by $\omega
\mapsto df \wedge \omega$ \\
$\di (X): {\mathcal{A}}^{p} \rightarrow {\mathcal{A}}^{p-1}$ by
contraction with the vectorfield $X$ \\
${\mathcal{L}}_{X} \equiv {\mathcal{L}}(X)=\di (X)\dd +\dd \di (X): {\mathcal{A}}^{p}
\rightarrow {\mathcal{A}}^{p}$ by the Lie derivative w.r.t. $X$

\item Representation of\\
fermionic creation operators: $\de(x^{m})$\\
fermionic annihilation operators: $\di(\partial/\partial x^{m})$

\item Operators on ${\mathcal{D}}_{\bullet} ({\mathbb{M}}^{D})$ \\
$M(f): {\mathcal{D}}_{p} \rightarrow {\mathcal{D}}_{p}$,
multiplication by scalar function $f : {\mathbb{M}}^{D} \rightarrow
{\mathbb{R}}$ \\
$\de (f): {\mathcal{D}}_{p} \rightarrow {\mathcal{D}}_{p-1}$ by
${\mathfrak{F}} \mapsto \dd f.{\mathfrak{F}}$ (contraction with the form $\dd f$)\\
$\di (X): {\mathcal{D}}_{p} \rightarrow {\mathcal{D}}_{p+1}$ by
multiplication and partial antisymmetrization \\
${\mathcal{L}}_{X} \equiv {\mathcal{L}}(X)=\di (X){\nabla}+{\nabla}\di (X): {\mathcal{D}}_{p}
\rightarrow {\mathcal{D}}_{p}$ by Lie derivative w.r.t. $X$

\item Forms and densities of weight one on $({\mathbb{M}}^{D}\,, g)$ \\
$C_{g} : {\mathcal{A}}^{p} \rightarrow {\mathcal{D}}_{p}$ (see eq.
21) \\
$\ast : {\mathcal{A}}^{p} \rightarrow {\mathcal{A}}^{D-p}$ s.t
${\mathcal{T}}(\omega |\eta) = \omega \wedge \ast \eta$ \\
$\delta : {\mathcal{A}}^{p+1} \rightarrow {\mathcal{A}}^{p}$ is the
metric transpose defined by \\
$~~~~~~[\dd\omega | \eta] =: [\omega | \delta\eta ] $ s.t.
$[\omega | \eta] := \int {\mathcal{T}} (\omega |\eta) $ \\
$~~~~~~\delta = C_{g}^{-1} b C_{g}~~~~~~~~~~~\mbox{(see eq. 28)} $ \\
$\beta : {\mathcal{D}}_{p+1} \rightarrow {\mathcal{D}}_{p}$ is
defined by $C_{g} \dd C_{g}^{-1}$

\item Gra\ss mann calculus on $\xi^{\lambda} \in \Lambda_{\nu}$ or
$\Lambda_{\infty}$ or unspecified $\Lambda$ \\
${\mathrm{d}\mathrm{d}} = 0$ remains true, therefore \\
$~~~~~~\frac{\partial}{\partial \xi^{\lambda}} \frac{\partial}{\partial
\xi^{\mu}} = -\frac{\partial}{\partial
\xi^{\mu}}\frac{\partial}{\partial\xi^{\lambda}} $ \\
$~~~~~~\dd \xi^{\lambda}\wedge\dd \xi^{\mu} = \dd \xi^{\mu}\wedge\dd \xi^{\lambda} $

\end{itemize}

\begin{itemize}
\item Forms and densities of weight $-1$ on ${\mathbb{R}}^{0|\nu}$ \\
Forms are graded totally symmetric covariant tensors. Densities are
graded totally symmetric contravariant tensors of weight -1. \\
$\left( {\mathcal{A}}^{\bullet}({\mathbb{R}}^{0|\nu})\,, \dd \right)$ Ascending
complex of forms not limited above \\
$\left( {\mathcal{D}}_{\bullet} ({\mathbb{R}}^{0|\nu})\,, \nabla\cdot \mbox{
or }{\mathrm{b}}\right)$ Descending complex of densities not limited above

\item Operators on ${\mathcal{A}}^{\bullet} ({\mathbb{R}}^{0|\nu})$ \\
$M(\varphi): {\mathcal{A}}^{p}({\mathbb{R}}^{0|\nu}) \rightarrow
{\mathcal{A}}^{p} ({\mathbb{R}}^{0|\nu})$ multiplication by a scalar function
$\varphi$ \\
$\de (\varphi): {\mathcal{A}}^{p} ({\mathbb{R}}^{0|\nu}) \rightarrow
{\mathcal{A}}^{p+1} ({\mathbb{R}}^{0|\nu})$ by $\de (\varphi) = \dd\varphi
\wedge$ \\
$\di (\Xi) : {\mathcal{A}}^{p} ({\mathbb{R}}^{0|\nu}) \rightarrow
{\mathcal{A}}^{p-1} ({\mathbb{R}}^{0|\nu})$ by contraction with the vectorfield $\Xi$ \\
${\mathcal{L}}_{\Xi} \equiv {\mathcal{L}}(\Xi) := \di (\Xi) \dd - {\mathrm{di}}(\Xi)$
maps ${\mathcal{A}}^{p}({\mathbb{R}}^{0|\nu}) \rightarrow {\mathcal{A}}^{p}
({\mathbb{R}}^{0|\nu})$

\item Representation of\\
bosonic creation operators: $\de(\xi^{\mu})$\\
bosonic annihilation operators: $\di(\partial/\partial\xi^{\mu})$

\item Operators on ${\mathcal{D}}_{\bullet}({\mathbb{R}}^{0|\nu})$ \\
$M(\varphi) : {\mathcal{D}}_{p} ({\mathbb{R}}^{0|\nu}) \rightarrow
{\mathcal{D}}_{p} ({\mathbb{R}}^{0|\nu})$, multiplication by scalar function
$\varphi$ \\
$\de (\varphi) : {\mathcal{D}}_{p} ({\mathbb{R}}^{0|\nu}) \rightarrow
{\mathcal{D}}_{p-1} ({\mathbb{R}}^{0|\nu})$ by ${\mathfrak{F}} \mapsto 
\dd (\varphi)\cdot {\mathfrak{F}}$ \\
$\di (\Xi) : {\mathcal{D}}_{p} ({\mathbb{R}}^{0|\nu}) \rightarrow
{\mathcal{D}}_{p+1} ({\mathbb{R}}^{0|\nu})$ by multiplication and partial symmetrization \\
${\mathcal{L}}_{\Xi} \equiv {\mathcal{L}}(\Xi)=\di (\Xi){\nabla}-{\nabla}\di (\Xi) : {\mathcal{D}}_{p}
({\mathbb{R}}^{0|\nu}) \rightarrow {\mathcal{D}}_{p}({\mathbb{R}}^{0|\nu})$ by Lie
derivative w.r.t. $\Xi$
\end{itemize}

\section{Recalling classical results}
\setcounter{equation}{0}

\subsection{Forms and densities on a $D$-dimensional manifold}

\indent We single out the statements which are independent of the dimension
$D$
of the manifold because our ultimate goal is integration on infinite
dimensional spaces.

We begin with properties of forms and densities which can be
established in the absence of a metric tensor because they are readily
useful in Gra\ss mann calculus; we then consider forms
and densities on riemannian manifolds $({\mathbb{M}}^{D}\,, g)$.

By ``forms'' we mean exterior differential forms, i.e.~totally
antisymmetric covariant tensors.

By ``densities'' we mean tensor-densities of weight one, i.e.~totally
antisymmetric contravariant tensors of weight one.  A density
${\mathfrak{F}}$ is said to be of weight $w$ if, under the change of
coordinate $\bar{x}(x)$, the new density $\bar{\mathfrak{F}}$ is
proportional to $(\det \partial\bar{x}/\partial x)^{w}$.

\subsection{Forms}

The exterior differentiation $\dd$ on the graded algebra
${\mathcal{A}}^{\bullet}$
of forms is a derivation\,\footnote{~The distinction between derivation and
anti-derivation according to the operator parity is not necessary if
one uses the graded Leibnitz rule.}
of degree 1.  Let ${\mathcal{A}}^{p}$ be the space of $p$-forms on
${\mathbb{M}}^{D}$,
\begin{equation}
   \dd: {\mathcal{A}}^{p} \longrightarrow {\mathcal{A}}^{p+1}
\end{equation}
In coordinates, using the convention that capitalized indices are
ordered,
\begin{eqnarray}
   \dd\omega &=& \frac{\partial \omega_{I_{1}\ldots I_{p}}}{\partial
   x^{m}} \dd x^{m} \wedge \dd x^{I_{1}} \wedge \ldots \wedge
   \dd x^{I_{p}}\nonumber \\
   &=& \varepsilon\begin{array}{l} m I_{1} \ldots I_{p}
   \\[-.02in] J_{1} \ldots
   J_{p+1} \end{array} \frac{\partial \omega_{I_{1}\ldots I_{p}}}{\partial
   x^m} \dd x^{J_{1}}\wedge \ldots \wedge \dd x^{J_{p+1}}\nonumber \\
   &=& \frac{1}{(p+1)!} \theta_{j_{1}\ldots j_{p+1}} \dd x^{j_{1}} \wedge
   \ldots \wedge \dd x^{j_{p+1}}\nonumber
\end{eqnarray}
which defines the components of $\theta = \dd\omega$.  We recall the
following operators on ${\mathcal{A}}^{\bullet}$ and some of their properties
\begin{equation}
   \begin{array}{lcll}
      M(f) & : & {\mathcal{A}}^{j} \longrightarrow {\mathcal{A}}^{j} &
      \mbox{by multiplication with } f:{\mathbb{M}}^{D}\rightarrow
      {\mathbb{R}}  \\
      \de (f) & : & {\mathcal{A}}^{j} \longrightarrow {\mathcal{A}}^{j+1} &
      \mbox{by }\omega \mapsto df \wedge \omega \\
      \di (X) & : & {\mathcal{A}}^{j} \longrightarrow {\mathcal{A}}^{j-1} &
      \mbox{by contraction with the vector } X \\
      {\mathcal{L}}_{X} \equiv {\mathcal{L}}(X) & : & {\mathcal{A}}^{j}
      \longrightarrow {\mathcal{A}}^{j} & \mbox{by the Lie derivative
w.r.t. } X
   \end{array}
\end{equation}
Their graded commutators are
\begin{eqnarray}
   &  & \left[ \de (f)\,, \de (g) \right]_{+} = 0 \\
   &  & \left[ \di (X)\,, \di (Y) \right]_{+} = 0 \\
   &  & \left[ \di (X)\,, \de (f) \right]_{+} = M({\mathcal{L}}_{X} f)
\end{eqnarray}
With respect to the exterior differential $\dd$ we have the following
graded commutators
\begin{eqnarray}
   \left[ \dd\,, \de (f) \right]_{+} &=& 0 \\
   \left[ \dd\,, \di (X) \right]_{+} &=& {\mathcal{L}}_{X} \\
   \left[ \dd\,, {\mathcal{L}}_{X} \right]_{-} &=& 0
\end{eqnarray}
which may be obtained from the explicit representation
\begin{equation}
   \dd = \de(x^{m}) {\mathcal{L}}(\partial/\partial x^{m})\,.
\end{equation}
Interpreting the degree $p$ of a form as a particle number, we note
that $\de(x^{m})$ is a {\it fermionic} creation operator and
$\di(\partial/\partial x^{m})$ is a {\it fermionic} annihilation operator.

\subsection{Densities}

A $p$-density is a density whose contraction with a $p$-form is a
scalar-density. In the coordinate basis $(\partial/\partial x^j)_j$ 
dual to $(\dd x^j)_j$ and using capitalized (ordered) indices
\begin{equation}
{\mathfrak{F}}=\sqrt{{\mathrm{det}}\,g}\frac{\partial}{\partial 
x^{I_1}}\wedge...\wedge\frac{\partial}{\partial x^{I_p}}{\mathfrak{F}}^{I_1...I_p}
\end{equation}
A scalar density is simply ${\mathfrak{F}}$.
\par Let ${\mathcal{D}}_{p}$ be the space of
$p$-densities\,\footnote{~Densities cannot be multiplied, therefore
${\mathcal{D}}_{\bullet}$ is not a graded algebra, because the
antisymmetrized product of two densities of weight one is a density
of weight two.}, the divergence
operator
\begin{equation}
   \nabla\cdot : {\mathcal{D}}_{p} \longrightarrow
   {\mathcal{D}}_{p-1}\,.
\end{equation}
In coordinates
\begin{equation}\label{54}
   ( \nabla \cdot {\mathfrak{F}} )^{\nu\ldots} =
   \frac{\partial}{\partial x^{\mu}} {\mathfrak{F}}^{\mu\nu\ldots}
\end{equation}
To conform with standard practice for descending complexes, we shall write the divergence
operation $\nabla\cdot$ on densities also as ${\mathrm{b}}$ (for
boundary) typically using ${\mathrm{b}}$ in the context of complexes and $\nabla$ in the 
context of computation.\cite{Bourbaki}
We introduce the following operators on ${\mathcal{D}}_{\bullet}$
\begin{equation}
   \begin{array}{ll}
      M(f) : {\mathcal{D}}_{p}\longrightarrow {\mathcal{D}}_{p} &
      \mbox{by multiplication with }f: {\mathbb{M}}^{D}
      \rightarrow {\mathbb{R}} \\
      \de (f) : {\mathcal{D}}_{p} \longrightarrow {\mathcal{D}}_{p-1} &
      \mbox{by contraction with } \dd (f) \\
      \di (X) : {\mathcal{D}}_{p} \longrightarrow {\mathcal{D}}_{p+1} &
      \mbox{by multiplication and } \\
      & \mbox{partial antisymmetrization} \\
      {\mathcal{L}}_{X} \equiv {\mathcal{L}}(X) : {\mathcal{D}}_{p}
      \longrightarrow
      {\mathcal{D}}_{p} & \mbox{by the Lie derivative w.r.t. }X
   \end{array}
\end{equation}
{\it Example}:  Let ${\mathfrak{F}}$ be a 2-density, then
\begin{equation}\label{56}
   \left( \di (X) {\mathfrak{F}} \right)^{\alpha\beta\gamma} = X^{\alpha}
   {\mathfrak{F}}^{\beta\gamma} + X^{\beta}
   {\mathfrak{F}}^{\gamma\alpha} + X^{\gamma}
   {\mathfrak{F}}^{\alpha\beta}\,.
\end{equation}
As in the case of forms, we obtain the commutator 
relations:
\begin{eqnarray}
   &  & [\de (f)\,, \de (g)]_{+} = 0 \\
   &  & [\di (X)\,, \di (Y)]_{+} = 0 \\
   &  & [\de (f)\,, \di (X)]_{+} = M({\mathcal{L}}_X f)
\end{eqnarray}
and
\begin{eqnarray}
   &  & [\nabla\,, \de (f)]_{+} =0 \\
   &  & [\nabla\,, \di (X)]_{+} = \frac{\partial }{\partial x^\mu} X^\mu\\
   &  & [\nabla\,, {\mathcal{L}}_{X}]_{-}=0.
\end{eqnarray}
which may be obtained from the explicit representation\,\footnote{
~Given 
${\mathcal{L}}(X)=\di (X)\nabla+\nabla\di (X)$ and 
using the definition (\ref{54}) and the example (\ref{56}), ${\mathcal{L}}(\frac{\partial}{\partial x^m})=\frac{\partial}{\partial 
x^m}$. The final result follows by contracting with 
$\nabla(x^m)$.}
\begin{equation}
\nabla=\de (x^m){\mathcal{L}}(\partial/\partial x^m).
\end{equation}
Interpreting the degree $p$ of a density as a particle number 
(i.e. the sum of all occupation numbers), 
we note that $\de (x^m)$ is a {\it fermionic} annihilation 
operator and $\di (\partial/\partial x^m)$ is a {\it 
fermionic} creation operator.

\subsection{Ascending and descending complexes on ${\mathbb{M}}^{D}$}

\indent Since $\dd\dd=0$, the graded algebra ${\mathcal{A}}^{\bullet}$ is an
{\it ascending
complex} w.r.t. to the operator $\dd$
\begin{equation}
   {\mathcal{A}}^{0}\stackrel{\dd}{\longrightarrow}
   {\mathcal{A}}^{1} \longrightarrow \cdots \longrightarrow
   {\mathcal{A}}^{D}\,.
\end{equation}
Since ${\mathrm{b}}{\mathrm{b}}=0$, the following sequence is a {\it descending complex}.
\begin{equation}
   {\mathcal{D}}_{0} \stackrel{\mathrm{b}}{\longleftarrow} {\mathcal{D}}_{1}
   \stackrel{\mathrm{b}}{\longleftarrow} \cdots \longleftarrow {\mathcal{D}}_{D-1}
   \stackrel{\mathrm{b}}{\longleftarrow} {\mathcal{D}}_{D}
\end{equation}
Writing ${\mathcal{D}}^{-p}$ instead of ${\mathcal{D}}_{p}$ is standard
practise in
homological algebra, and the descending complex can be written as an
{\it ascending complex in negative degrees}
\begin{equation}
   {\mathcal{D}}^{-D} \stackrel{\mathrm{b}}{\longrightarrow}
   {\mathcal{D}}^{-D+1} \stackrel{\mathrm{b}}{\longrightarrow} \cdots
   \longrightarrow {\mathcal{D}}^{-1}\stackrel{\mathrm{b}}{\longrightarrow}
   {\mathcal{D}}^{0}\,.
\end{equation}
{\it Remark}: The operator $\di (X)$ moves downwards on the ascending
complex ${\mathcal{A}}^{\bullet}$, and upwards on the descending
complex ${\mathcal{D}}_{\bullet}$.

\subsection{Forms and densities on a riemannian manifold
$({\mathbb{M}}^{D}\,,g)$}\label{sec7.5}

\indent The metric tensor $g$ provides a correspondence $C_{g}$ between a
$p$-form and a $p$-density.  For instance, let $F$ by a 2 form, then
\begin{equation}
   {\mathfrak{F}}^{\alpha\beta} = \sqrt{\det g_{\mu\nu}} F_{\gamma\delta}
   g^{\alpha\gamma} g^{\beta\delta}
\end{equation}
are components of a 2-density.  The metric $g$ is used twice: a)
raising indices, b) introducing weight 1 by multiplication with
$\sqrt{\det g}$.  This correspondence does not depend on the
dimension $D$.

On an orientable manifold, the dimension $D$ can be used for
transforming a $p$-density into a $(D-p)$-form.  For example let $D=4$
and $p=1$
\begin{equation}
   t_{\alpha\beta\gamma} :=
   \varepsilon^{1234}_{\alpha\beta\gamma\delta} {\mathfrak{F}}^{\delta}
\end{equation}
where the alternating symbol $\varepsilon$ defines an orientation.

The star operator (Hodge-de Rham operator, see Ref.~\protect\citebk{Choquet-Bruhat}, p.~295)) combines the metric-dependent and the dimension-dependent
transformations; it transforms a $p$-form into a $(D-p)$-form by
\begin{equation}
   {\mathcal{T}} (\omega |\eta) = \omega \wedge \ast \eta
\end{equation}
where, as usual, the scalar product of 2 $p$-form $\omega$ and $\eta$
is
\begin{equation}
   (\omega|\eta) = \frac{1}{p!} \omega_{i_{1}\ldots i_{p}}
   \eta^{i_{1}\ldots i_{p}}
\end{equation}
and ${\mathcal{T}}$ is the volume element, given in example 1 below.

We shall exploit the correspondence mentioned in the first paragraph
\begin{equation}
   C_{g} : {\mathcal{A}}^{p} \longrightarrow {\mathcal{D}}_{p}
\end{equation}
for constructing a descending complex on ${\mathcal{A}}^{\bullet}$ w.r.t.
to the metric transpose $\delta$ of $\mathrm{d}$ (Ref.~\protect\citebk{Choquet-Bruhat}, p.~296)
\begin{equation}
   \delta : {\mathcal{A}}^{p+1} \longrightarrow {\mathcal{A}}^{p}
\end{equation}
and an ascending complex on ${\mathcal{D}}_{\bullet}$
\begin{equation}
   \beta : {\mathcal{D}}_{p} \longrightarrow {\mathcal{D}}_{p+1}
\end{equation}
where $\beta$ is defined by the following diagram
\begin{equation}
   \left.
   \begin{array}{clclc}
      ~ & {\mathcal{A}}^{p} & \begin{array}{c}
      \stackrel{\scriptstyle{\delta}}{\textstyle{\longleftarrow}}
      \\[-.05in] \stackrel{\textstyle{\longrightarrow}}{\scriptstyle{d}}
      \end{array} & {\mathcal{A}}^{p+1} & ~ \\[-.05in]
      C_{g} & \downarrow & ~ & \downarrow & C_{g} \\[-.05in]
      ~ & {\mathcal{D}}_{p} & \begin{array}{c}
      \stackrel{\beta}{\longrightarrow} \\[-.05in]
      \stackrel{\textstyle{\longleftarrow}}{\scriptstyle{b}} \end{array} &
      {\mathcal{D}}_{p+1} &~
  \end{array} \right\}  \Longleftrightarrow \left\{
   \begin{array}{rcl}
      \delta &=& C^{-1}_{g} b C_{g} \\[.12in] \beta &=& C_{g} d C^{-1}_{g}
   \end{array} \right.\,.
\end{equation}
\textit{Example 1}:  Volume element on an oriented $D$-dimensional
riemannian manifold.
The volume element is
\begin{equation}
{\mathcal{T}} := \dd x^{1} \wedge \ldots \wedge 
\dd x^{D}\,{\mathfrak{T}} =~ \dd x^{1} \wedge \ldots \wedge 
\dd x^{D}\,\sqrt{\det g}\mbox{ with }{\mathfrak{T}} \in {\mathcal{D}}_0
\end{equation}
where ${\mathfrak{T}}$ is a scalar density corresponding to the top form
\begin{equation}
   \dd x^{1}\wedge \ldots \wedge \dd x^{D}\in {\mathcal{A}}^D
\end{equation}
${\mathfrak{T}}$ is indeed a scalar density since, under the change of
coordinates $x^{\prime j} = A^{j}{}_{i} x^{i}$
\begin{equation}
   {\mathfrak{T}}^{\prime} = (\det A){\mathfrak{T}}\,.
\end{equation}
\textit{Example 2:} In the thirties, the use of densities was often justified 
by the fact that in a number of useful examples it reduces
the number of indices.  For example, a vector-density in
${\mathbb{M}}^{4}$ can replace a 3-form
\begin{equation}
   {\mathfrak{T}}^{l} = \sqrt{\det g}~
   \varepsilon^{ijkl}_{1234} t_{ijk}\,.
\end{equation}
An axial vector in ${\mathbb{R}}^{3}$ can replace a 2-form.

\subsection{Other definitions of the metric transpose $\delta$\,.}

\begin{itemize}
\item The name ``metric transpose of the differential $\dd$'' comes from
the integrated version of ${\mathcal{T}} (\omega|\eta) = \omega\wedge
\ast \eta$; namely let
\begin{equation}
   [ \omega|\eta ] = \int {\mathcal{T}} (\omega|\eta)
\end{equation}
then for $\omega$ a $p$-form and $\eta$ a $(p+1)$-form on a manifold
without boundary $\delta$ is defined by
\begin{equation}
   [\dd\omega |\eta ] =: [\omega |\delta\eta ]\,.
\end{equation}

\item On a $p$-form $\omega$,
\begin{equation}
   \delta\omega := (-1)^{p} \ast^{-1} \dd\ast \omega
\end{equation}
where $\ast$ is the star operator defined above.

\item $\delta$ is a derivation on ${\mathcal{A}}^{\bullet}$ of degree
-1.
\end{itemize}
We have given a new presentation of these well known results, but
we have separated metric-dependent and dimension-dependent
transformations.  It brings forth the ascending complex on densities
w.r.t. to the operator $\beta = C_{g} \dd C_{g}^{-1}$.

\section{Berezin integration}
\setcounter{equation}{0}

\subsection{A Berezin integral is a derivation}

The fundamental requirement on a definite integral is expressed in
terms of an integral operator $I$ and a derivative operator $D$ on a
space of functions, namely
\begin{equation}
   DI = ID = 0
\end{equation}

The requirement $DI=0$ for functions of real variables $f:{\mathbb{R}}^{D}
\longrightarrow {\mathbb{R}}$ says that the integral does not depend on the
variable of integration
\begin{equation}
   \frac\dd {\dd x} \int f(x) \dd x = 0\,, \qquad x \in {\mathbb{R}}\,.
\end{equation}

The requirement $ID=0$ on the space of functions defined on domains
with vanishing boundaries says
\begin{equation}
   \int \frac{\dd}{\dd x} f(x) \dd x = 0
\end{equation}
This is the foundation of integration by parts
\begin{equation}
   0 = \int \dd\left( f(x)g(x) \right) = \int \dd f(x) \cdot g(x) + \int
   f(x) \dd g(x)\,,
\end{equation}
and of the Stokes' theorem on a form $\omega$,
\begin{equation}
   \int_{\mathbb{M}} \dd\omega = \int_{\partial\mathbb{M}}\omega = 0
   \qquad \mbox{since } \partial \mathbb{M} \mbox{ is an empty set}
\end{equation}
We shall use the requirement $ID = 0$ in section II.5 for imposing a
condition on volume elements.

We now use the fundamental requirements on Berezin integrals defined on
functions $f$ of the Grassman algebra $\Lambda_{\nu}$.  The condition
$DI=0$ says
\begin{equation}
   \frac{\partial}{\partial\xi^{i}} I(f) = 0\qquad \qquad \mbox{for }
   i\in \{1\,, \ldots\,, \nu\}
\end{equation}
Any operator on $\Lambda_{\nu}$ can be set in normal
ordering\,\footnote{~This ordering is also the operator normal ordering,
creation operator followed by annihilation operator, since $\de (\xi^\mu)$ and
$\di (\partial/\partial \xi^\mu)$ can be interpreted as creation 
and annihilation operators (see (\ref{97}) to (\ref{99})).}
\begin{equation}
   \Sigma C_{K}{}^{J} \xi^{K} \frac{\partial}{\partial \xi^{J}}
\end{equation}
with $J\,,K\,,$ multi-ordered indices.  Therefore the condition
$DI=0$ implies that $I$ is a polynomial in $\partial/\partial
\xi^{i}$,
\begin{equation}
   I = Q \left( \frac{\partial}{\partial\xi^{1}}\,, \ldots
   \frac{\partial}{\partial\xi^{\nu}}\right)\,.
\end{equation}
The condition $ID = 0$, namely
\begin{equation}
   Q \left( \frac{\partial}{\partial \xi^{1}}\,, \ldots\,,
   \frac{\partial}{\xi^{\nu}} \right)
   \frac{\partial}{\partial\xi^{\mu}} = 0 \qquad \mbox{for every }
   i\in \{ 1\,, \ldots\,, \nu \}\,,
\end{equation}
implies
\begin{equation}
   I = \mbox{constant } \frac{\partial}{\partial\xi^{\nu}} \cdots
   \frac{\partial}{\partial\xi^{1}}
\end{equation}
A Berezin integral is a derivation.  The constant is a normalisation
constant chosen for convenience in the given context.  Usual choices
include $1$, $(2\pi i)^{1/2}$, $(2\pi i)^{-1/2}$.

\subsection{Change of variable of integration}

Since integrating $f(\xi^{1}\,, \ldots\,, \xi^{\nu})$ is taking its
derivatives w.r.t. $\xi^{1}\,, \ldots\,, \xi^{\nu}$, a change
of variable of integration is most easily performed on the
derivatives.  Given a change of coordinates $f$, we recall the induced
transformations on the tangent and cotangent spaces.  Let $y=f(x)$
and $\theta = f(\zeta)$;
\begin{figure}[h]
\centerline{\includegraphics[width=12cm,totalheight=4.5cm]
{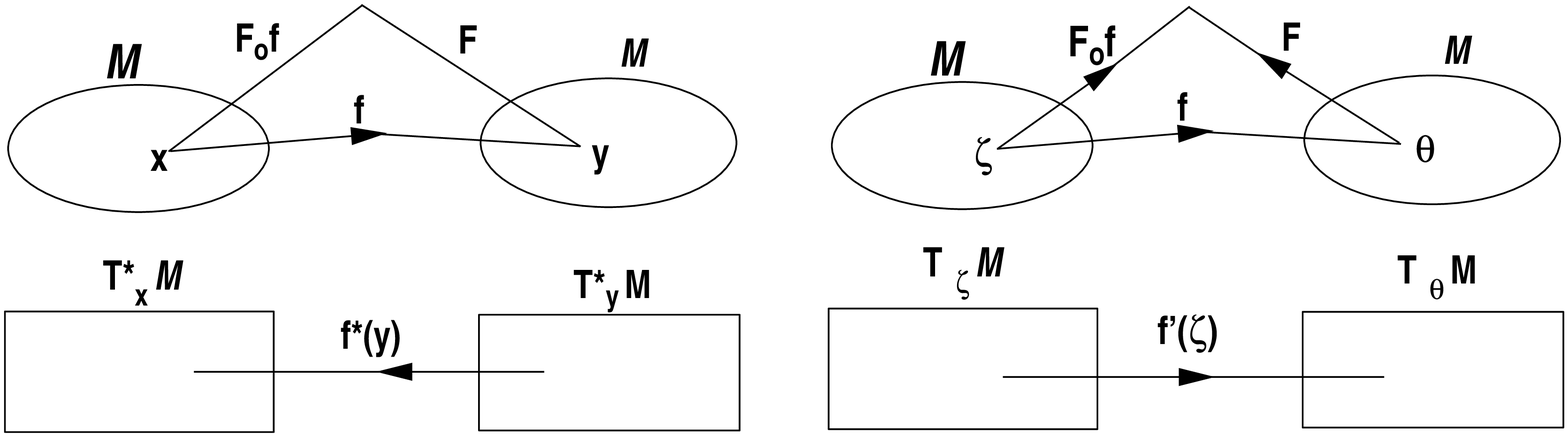}}
\end{figure}
\begin{equation}
   \dd y^{1} \wedge \ldots \wedge \dd y^{D} =  \dd x^{1} \wedge \ldots
   \wedge \dd x^{D}\left( \det
   \frac{\partial y^{i}}{\partial x^{j}} \right)
\end{equation}
and
\begin{equation}\label{87}
   \int \dd x^{1} \wedge \ldots \wedge \dd x^{D}\,(F \circ f)(x) \left(\det \frac{\partial f^{i}}{\partial
   x^{j}} \right)= \int \dd y^{1}
   \wedge \ldots \wedge \dd y^{D}\,F(y)
\end{equation}
On the other hand, for an intregral over Gra\ss mann variables, the
antisymmetry leading to a determinant is the antisymmetry of the
product $\partial_{1}\ldots \partial_{D}$.  And
\begin{equation}\label{88}
   \left( \frac{\partial}{\partial\zeta^{1}} \ldots
   \frac{\partial}{\partial \zeta^{D}} \right) (F \circ f)(\zeta) =
   \left( \det \frac{\partial\theta^{\lambda}}{\partial\zeta^{\mu}} \right)
   \frac{\partial}{\partial\theta^{1}}\cdots
   \frac{\partial}{\partial\theta^{D}} F(\theta)
\end{equation}
The determinant is now on the right hand side, it will become an
inverse determinant when brought to the same side as in 
(\ref{87}).

\section{Ascending and descending complexes in Gra\ss mann calculus}
\setcounter{equation}{0}

In section \ref{sec7.5} we have presented $\dd$-ascending and
$\delta$-descending complexes ${\mathcal{A}}^{\bullet}({\mathbb{M}})$
of forms and ${\mathrm{b}}$-descending and $\beta$-ascending complexes
${\mathcal{D}}_{\bullet}({\mathbb{M}})$ of densities.  We shall now
study complexes ${\mathcal{A}}^{\bullet} ({\mathbb{R}}^{0 |\nu})$ of Gra\ss mann
forms, and complexes ${\mathcal{D}}_{\bullet} ({\mathbb{R}}^{0 |\nu})$ of
Gra\ss mann densities.

\subsection{Gra\ss mann Forms}

Two properties of forms on real variables remain true for forms on
Gra\ss mann variables, namely
\begin{equation}
   \dd\dd\omega = 0
\end{equation}
\begin{equation}
   \dd(\omega \wedge \theta) = \dd\omega \wedge \theta +
   (-1)^{\tilde{\omega}\tilde{d}} \omega \wedge d\theta\,;
\end{equation}
a form on Gra\ss mann variables is a {\it graded} totally antisymmetric
covariant tensor.  Indeed
\begin{equation}
   \xi^{\lambda} \xi^{\mu} = -\xi^{\mu}\xi^{\lambda}
\end{equation}
implies
\begin{equation}
   \partial_{\lambda} \partial_{\mu} =
   -\partial_{\mu}\partial_{\lambda}\qquad
   \qquad \mbox{where} \qquad \partial_{\lambda} := \partial/\partial
   \xi^{\lambda}
\end{equation}
which, in turn implies
\begin{equation}\label{93}
   d\xi^{\lambda}\wedge d\xi^{\mu} = d\xi^{\mu} \wedge 
   d\xi^{\lambda}.
\end{equation}
The counterparts on ${\mathcal{A}}^{\bullet} ({\mathbb{R}}^{0|\nu})$ of the
operators $M(f)$, $\de (f)$, $\di (X)$, ${\mathcal{L}}(X)$ on
${\mathcal{A}}^{\bullet}({\mathbb{M}})$ are as follows; we omit the
reference to ${\mathbb{R}}^{0|\nu}$ for visual clarity.
\begin{equation}
   \begin{array}{rcll}
      M(\varphi) & : & {\mathcal{A}}^{j} \longrightarrow
      {\mathcal{A}}^{j} & ~~~~\mbox{by multiplication by } \varphi:
      {\mathbb{R}}^{0|\nu} \rightarrow \Lambda_{\nu} \\
      \de (\varphi) & : & {\mathcal{A}}^{j} \longrightarrow
      {\mathcal{A}}^{j+1} & ~~~~\mbox{by } \de (\varphi) :=
      \dd\varphi\wedge \\
      \di (\Xi) & : & {\mathcal{A}}^{j} \longrightarrow
      {\mathcal{A}}^{j-1} & ~~~~\mbox{by contraction with } \Xi\\
      {\mathcal{L}}_{\Xi} \equiv {\mathcal{L}}(\Xi) & : &
      {\mathcal{A}}^{j} \longrightarrow {\mathcal{A}}^{j} &
      ~~~~\mbox{by Lie derivative with respect to }\Xi
    \end{array}
\end{equation}
We note the following properties\,\footnote{
~The difference to ordinary forms is due to the symmetrization in 
the case of Gra\ss mann forms
\begin{equation}
(\omega\wedge\pi)(\Xi_1,...,\Xi_{r+s})=
\frac{1}{r!s!}\sum_{P(1..r+s)}\omega(\Xi_{P(1)},...\Xi_{P(r)})\pi(\Xi_{P(r+1)},...\Xi_{P(r+s)})\nonumber
\end{equation}
contrary to the antisymmetrization in the case of ordinary forms
\begin{equation}
(\omega\wedge\pi)(X_1,...,X_{r+s})=
\frac{1}{r!s!}\sum_{P(1..r+s)}\mbox{sgn}(P)\omega(X_{P(1)},...X_{P(r)})\pi(X_{P(r+1)},...X_{P(r+s)})\nonumber.
\end{equation}
}
\begin{equation}
   \di (\Xi) (\omega^{k}\wedge \eta) = \left( \di (\Xi) \omega^{k}\right)
   \wedge \eta + \omega^{k} \wedge \di (\Xi) \eta \qquad
   \omega^{k} \in {\mathcal{A}}^{k}
\end{equation}
i.e.~the parity of $\di (\Xi)$ is zero.  It follows that
\begin{equation}
   {\mathcal{L}}(\Xi) = \di (\Xi)d - d\di (\Xi)
\end{equation}
For $\varphi$, $\xi$, $\Xi$, $\Psi$ odd, the corresponding graded
commutators are
\begin{eqnarray}
   \left[ \de (\varphi)\,, \de (\zeta) \right]_{-} &=& 0 \label{97}\\
   \left[ \di (\Xi)\,, \di (\Psi) \right]_{-} &=& 0 \\
   \left[ \di (\Xi)\,, \de (\varphi) \right]_{-} &=& 
   M({\mathcal{L}}_{\Xi}\label{99}
   \varphi) \\
   \left[ \dd\,, \de (\varphi) \right]_{-} &=& 0 \\
   \left[ \dd\,, \di (\Xi) \right]_{-} &=& {\mathcal{L}}_{\Xi} \\
   \left[ \dd\,, {\mathcal{L}} (\Xi) \right]_{+} &=& 0
\end{eqnarray}
which may be obtained from the explicit representation
\begin{equation}
   \dd = \de(\xi^{\mu}) {\mathcal{L}} \left( \partial/\partial \xi^{\mu}
   \right)
\end{equation}
Interpreting the degree $p$ of a Gra\ss mann form as a particle number,
we note that $\de(\xi^{\mu})$ is a {\it bosonic} creation operator and
$\di(\partial/\partial\xi^{\mu})$ is a {\it bosonic} annihilation
operator.
\par Since the differential of a Gra\ss mann variable has even parity 
(\ref{93}), there is at first no reason to restrict our 
forms to polynomials in $\mathrm\dd \xi^i$. (The space of polynomials
is a proper subset of the space of smooth functions.) This leads to arbitrary 
smooth functions $\omega=\omega(\xi^i,\mathrm\dd \xi^i)$, which 
Voronov\,\cite{Voronov} calls {\it pseudodifferential forms}. Those forms 
can obviously no longer be decomposed in even and odd parts. Since 
we consider this necessary for the description of the quantum Fock 
space, we restrict our considerations to polynomial forms.

\subsection{Gra\ss mann Densities}

In order to define Gra\ss mann densities we recall the definitions of
densities in the works of H.~Weyl\,\cite{Weyl} (1920) W.~Pauli\,\cite{Pauli} (1921) and L.~Brillouin\,\cite{Brillouin} (1938).  From Pauli's
{\it Theory of Relativity} (p. 32):
``If the integral $\int{\mathfrak{F}}\dd x$ is an invariant (in a change of
coordinate system), then ${\mathfrak{F}}$ is called a scalar density,
following Weyl's terminology'' (in {\it Space-Time-Matter} pp 109
ff).  Here $\dd x$ stands for $\dd x^{1} \wedge \ldots \wedge \dd x^{D}$.
Under the change of variable $y(x)$, ${\mathfrak{F}}$ is a scalar density if
it is a scalar multiplied by $\det (\partial y^{j}/\partial 
x^{i})$.

What we call here ``tensor-densities'' is called ``linear tensor
densities'' by Weyl.  For him the term ``tensor densities'' are
arbitrary tensors of weight one.  He singles out among them the
contravariant antisymmetric\,\footnote{~The word ``skew'' is missing in the
English translation.  The original reads ``Die gleiche ausgezeichnete
Rolle, welche unter den Tensoren die kovarianten schiefsymmetrischen
spielen, kommt unter den Tensordichten den kontravarianten
schiefsymmetrischen zu, die wir darum kurz als {\it lineare
Tensordichten} bezeichnen wollen.''} ones and calls them
``linear'' because, like the ``covariant skew-symmetrical tensors''
(i.e.~the forms), they play a ``unique part.''  Their unique
properties, algebraic and geometrical are beautifully presented in
Brillouin's book.  Brillouin defines ``capacity'' as
an object which multiplied by a density is a scalar.  Together
densities and capacities have become known as ``pseudo-tensors'' and are
sometimes treated as ``second rate'' tensors!

If the Berezin integral $\int \dd \xi^\nu...\dd \xi^1\,f(\xi^{1},\ldots ,\xi^{\nu})= \frac{\partial}{\partial\xi^{\nu}} \cdots
\frac{\partial}{\partial\xi^{1}} f(\xi^{1},\ldots ,\xi^{\nu})
$ is invariant under the change of
coordinates $\theta (\xi)$ then $f$ is a Gra\ss mann scalar density.
It follows from the formula for change of variable of integration
(\ref{88}) that a Gra\ss mann scalar density is a scalar divided by $\det
\left( \partial \theta^{\lambda}/\partial \xi^{\mu}\right)$.  The
expression $\frac{\partial}{\partial\xi^{\nu}} \cdots
\frac{\partial}{\partial\xi^{1}}$ is a Gra\ss mann capacity in
${\mathbb{R}}^{0|\nu}$.

Two properties of densities on real variables remain true for
Gra\ss mann densities ${\mathfrak{F}}$, namely
\begin{equation}
   (\nabla\cdot)(\nabla\cdot){\mathfrak{F}} = 0
\end{equation}
\begin{equation}
   (\nabla\cdot)(X{\mathfrak{F}}) = (\nabla\cdot X) \cdot
   {\mathfrak{F}} + (-1)^{\tilde{X}\tilde{\nabla}} X\nabla\cdot
   {\mathfrak{F}}
\end{equation}
where $X$ is a vector field.  The first property follows from the fact
that the Gra\ss mann divergence is odd and the density is a graded
antisymmetric contraviant tensor, i.e.~symmetric in the interchange of
two Gra\ss mann indices.

The second property is the ``Leibnitz'' property of divergence over
products.  A density is a tensor of weight 1; multiplication by a
tensor of weight zero is the only possible product which maps a
density into a density.

Together these two properties make possible the construction of a
density complex.

Multiplying a Gra\ss mann scalar density ${\mathfrak{F}}$ by a graded totally
antisymmetric contravariant tensor gives a Gra\ss mann tensor density of
components
\begin{equation}
   {\mathfrak{F}}^{\mu\nu\rho\ldots}
\end{equation}
The counterparts on ${\mathcal{D}}_{\bullet}({\mathbb{R}}^{0|\nu})$ of the
operators $M(f)$, $\de (f)$, $\di (X)$, ${\mathcal{L}}(X)$ on
${\mathcal{D}}_{\bullet}({\mathbb{M}})$ are as follows.  We omit the
reference to ${\mathbb{R}}^{0|\nu}$ for visual clarity.
\begin{equation}
\begin{array}{ll}
   M(\varphi) : {\mathcal{D}}_{p} \longrightarrow
   {\mathcal{D}}_{p} ~~&~~ \mbox{by multiplication by } \varphi :
   {\mathbb{R}}^{0|\nu} \longrightarrow {\mathbb{R}} \\
   \de (\varphi): {\mathcal{D}}_{p} \longrightarrow
   {\mathcal{D}}_{p-1} ~~&~~ \mbox{by contraction with }
   \dd (\varphi) \\
   \di (\Xi) : {\mathcal{D}}_{p} \longrightarrow
   {\mathcal{D}}_{p+1} ~~&~~ \mbox{by multiplication and} \\
   &~~ \mbox{partial antisymmetrization} \\
   {\mathcal{L}}_{\Xi} \equiv {\mathcal{L}}(\Xi): {\mathcal{D}}_{p}
   \longrightarrow {\mathcal{D}}_{p} ~~&~~ \mbox{by the Lie
   derivative w.r.t. } \Xi
\end{array}
\end{equation}

\subsection{Ascending and descending Gra\ss mann complexes}
The ascending complex ${\mathcal{A}}^\bullet({\mathbb{R}}^{0|\nu})$ 
of Gra\ss mann forms with respect to $\dd $ does not 
terminate at the $\nu$-form. The descending complex 
${\mathcal{D}}_\bullet({\mathbb{R}}^{0|\nu})$ of densities with 
respect to $\nabla$ does not terminate at the $\nu$-density. 
Indeed, whereas forms and densities on ordinary variables are 
antisymmetric tensors, on Gra\ss mann variables they are symmetric 
tensors, therefore their degrees are not limited to the Gra\ss 
mann dimension $\nu$.

\subsection{Summary of complexes on ordinary and Gra\ss mann 
variables}
On ${\mathbb{M}}^D$:
\begin{eqnarray}
{\mathcal{D}}_D\longrightarrow\phantom{{\mathcal{D}}_1\longrightarrow}&...&\longrightarrow{\mathcal{D}}_0\,\\
{\mathcal{A}}^0\,\stackrel{\dd}\longrightarrow{\mathcal{A}}^1\stackrel{\dd}\longrightarrow&...&\stackrel{\dd}\longrightarrow{\mathcal{A}}^D\\
{\mathcal{D}}_0\,\stackrel{\mathrm{b}}\longleftarrow{\mathcal{D}}_1\stackrel{\mathrm{b}}\longleftarrow&...&
\stackrel{\mathrm{b}}\longleftarrow{\mathcal{D}}_D
\end{eqnarray}
${\mathcal{D}}_0\rightarrow{\mathcal{A}}^D$ by a 
dimension-dependent equation (a scalar density is the strict 
component of a top form)
\\${\mathcal{A}}^0\rightarrow{\mathcal{D}}_0$ by a 
metric-dependent equation.
\begin{itemize}
\item on ${\mathcal{A}}^\bullet({\mathbb{M}}^D)$, 
\begin{tabular}{l l}
$\de (x^k)$ & represents a fermionic creation operator \\
$\di (\partial/\partial x^k)$ & represents a fermionic 
annihilation operator.
\end{tabular}
\item on ${\mathcal{D}}_\bullet({\mathbb{M}}^D)$, 
\begin{tabular}{l l}
$\de (x^k)$ & represents a fermionic annihilation operator \\
$\di (\partial/\partial x^k)$ & represents a fermionic 
creation operator.
\end{tabular}
\end{itemize}
On ${\mathbb{R}}^{0|\nu}$:
\begin{eqnarray}
...{\mathcal{D}}_\nu\longrightarrow\phantom{{\mathcal{D}}_1\longrightarrow}&...&\longrightarrow{\mathcal{D}}_0\\
\phantom{...}{\mathcal{A}}^0\,\stackrel{\dd}\longrightarrow{\mathcal{A}}^1\stackrel{\dd}\longrightarrow&...&\stackrel{\dd}\longrightarrow{\mathcal{A}}^\nu...\\
\phantom{...}{\mathcal{D}}_0\,\stackrel{\mathrm{b}}\longleftarrow{\mathcal{D}}_1\stackrel{\mathrm{b}}\longleftarrow&...&
\stackrel{\mathrm{b}}\longleftarrow{\mathcal{D}}_\nu...
\end{eqnarray}
\begin{itemize}
\item on ${\mathcal{A}}^\bullet({\mathbb{R}}^{0|\nu})$, 
\begin{tabular}{l l}
$\de (\xi^\mu)$ & represents a bosonic creation operator \\
$\di (\partial/\partial \xi^\mu)$ & represents a bosonic
annihilation operator.
\end{tabular}
\item on ${\mathcal{D}}_\bullet({\mathbb{R}}^{0|\nu})$, 
\begin{tabular}{l l}
$\de (\xi^\mu)$ & represents a bosonic annihilation operator \\
$\di (\partial/\partial \xi^\mu)$ & represents a bosonic 
creation operator.
\end{tabular}
\end{itemize}

\section{The mixed case}
\setcounter{equation}{0}

\subsection{Integration over ${\mathbb{R}}^{n|\nu}$}\label{10.1}
We consider superfunctions on ${\mathbb{R}}^{n|\nu}$, that is 
functions of $n$ real variables $x^a$ and $\nu$ Gra\ss mann 
variables $\xi^\alpha$. Such a superfunction is of the form
\begin{equation}
F(x,\xi)=\sum_{p=0}^\nu\frac{1}{p!}f_{\alpha_1...\alpha_p}(x)\xi^{\alpha_1}...\xi^{\alpha_p}
\end{equation}
where the functions $f_{\alpha_1...\alpha_p}$ are smooth functions 
on ${\mathbb{R}}^n$, antisymmetrical in the indices 
$\alpha_1,...,\alpha_p$.
\par By definition, the integral of $F(x,\xi)$ is obtained by 
integrating w.r.t. the real variables, and performing a Berezin 
integral over the Gra\ss mann variables:
\begin{equation}
\int_{{\mathbb{R}}^{n|\nu}}\dd (x,\xi)\,F(x,\xi):=\int_{{\mathbb{R}}^{n}}\dd x\,\left(
\int_{{\mathbb{R}}^{0|n}}\dd \xi\, F(x,\xi) \right).
\end{equation}
More explicitly
\begin{equation}
\int_{{\mathbb{R}}^{n|\nu}}\dd (x,\xi)\,F(x,\xi)=\int_{{\mathbb{R}}^{n}}\dd 
^nx\,f_{12...\nu}(x).
\end{equation}
$\dd ^nx=\dd x^1...\dd x^n$ as usual. A 
theorem of Fubini type holds:
\begin{equation}
\int \dd ({\mathbf{x}},{\mathbf{y}})\,
F({\mathbf{x}},{\mathbf{y}})
=\int\dd {\mathbf{x}}\,\left(\int\dd {\mathbf{y}}\, F({\mathbf{x}},{\mathbf{y}})\right)
\end{equation}
where ${\mathbf{x}}=(x^a,\xi^\alpha)$ runs over 
${\mathbb{R}}^{n|\nu}$ and ${\mathbf{y}}=(y^b,\eta^\beta)$ over
${\mathbb{R}}^{m|\mu}$, hence 
$({\mathbf{x}},{\mathbf{y}})=(x^a,y^b,\xi^\alpha,\eta^\beta)$ over 
${\mathbb{R}}^{n+m|\nu+\mu}$. In particular
\begin{equation}
\int \dd (x,\xi)\,
F(x,\xi)=\int_{{\mathbb{R}}^{0|\nu}}\dd \xi\,\int_{{\mathbb{R}}^n}\dd ^nx\,F(x,\xi).
\end{equation}

\subsection{Scalar densities over ${\mathbb{R}}^{n|\nu}$}
A {\it scalar density} over ${\mathbb{R}}^{n|\nu}$ is simply a 
superfunction $D(x,\xi)$ used for integration purposes
\begin{equation}
\int_{{\mathbb{R}}^{n|\nu}}\dd T\cdot F=\int_{{\mathbb{R}}^{n|\nu}}\dd 
(x,\xi)\,D(x,\xi)F(x,\xi)
\end{equation}
Explicitly, we expand $D(x,\xi)$ as 
\begin{equation}
D(x,\xi)=\sum_{p=0}^\nu\frac{1}{p!}D_{\alpha_1...\alpha_p}(x)\xi^{\alpha_1}...\xi^{\alpha_p}
\end{equation}
with antisymmetric coefficients $D_{\alpha_1...\alpha_p}$. Hence
\begin{equation}
\int_{{\mathbb{R}}^{n|\nu}}\dd T\cdot F=\int_{{\mathbb{R}}^n}\dd ^nx\,
G_{1...\nu}(x)
\end{equation}
where
\begin{equation}
G_{\alpha_1...\alpha_\nu}=\sum_{p=0}^\nu\left(
\begin{array}{c}
\nu \\
p \\
\end{array}
\right)
D_{[\alpha_1...\alpha_p}F_{\alpha_{p+1}...\alpha_\nu]}
\end{equation}
Using the totally antisymmetric symbol
$\epsilon^{\alpha_1...\alpha_\nu}$ normalized by 
$\epsilon^{1...\nu}=1$, we raise indices as follows
\begin{equation}
D^{\alpha_1...\alpha_p}=\frac{1}{(\nu-p)!}\epsilon^{\alpha_1...\alpha_p\alpha_{p+1}...\alpha_\nu}D_{\alpha_{p+1}...\alpha_\nu}
\end{equation}
Then we obtain
\begin{equation}
\int_{{\mathbb{R}}^{n|\nu}}\dd T\cdot F=\int_{{\mathbb{R}}^n}\dd ^nx\,
G(x)
\end{equation}
where
\begin{equation}
G(x)=\sum_{p=0}^\nu\frac{1}{p!}D^{\alpha_1...\alpha_p}(x)F_{\alpha_1...\alpha_p}(x)
\end{equation}
Introducing a differential operator $\Lambda$ acting on the Gra\ss 
mann variables
\begin{equation}
\Lambda=\sum_{p=0}^\nu\frac{1}{p!}D^{\alpha_1...\alpha_p}\frac{\partial}{\partial 
\xi^{\alpha_1}}...\frac{\partial}{\partial \xi^{\alpha_p}}
\end{equation}
(recall that the $\frac{\partial}{\partial \xi^\alpha}$ mutually 
anticommute), then we get
\begin{equation}
G(x)=(\Lambda F)(x,0)
\end{equation}
(putting the $\xi^\alpha=0$). Finally
\begin{equation}\label{10.14}
\int  
\dd T\cdot F=\int_{{\mathbb{R}}^n}\dd ^nx\,\Lambda F(x,\xi)\mid_{\xi=0};
\end{equation}
hence the {\it mixed integration is really an integrodifferential 
operator, differentiating w.r.t. the Gra\ss mann variables, integrating w.r.t. 
the ordinary variables}.

\subsection{An analogy}
Let us consider a real space ${\mathbb{R}}^{n+m}$ of $n+m$ 
dimensions and embed the $n$-space ${\mathbb{R}}^n$ in 
${\mathbb{R}}^{n+m}$ as the set of vectors 
$(x^1,...,x^n,0,...,0)$. A well-known result by Laurent Schwartz 
asserts that a distribution on ${\mathbb{R}}^{n+m}$ carried by the 
subspace ${\mathbb{R}}^n$ is a sum of transversal 
derivatives of distributions\,\footnote{~We use coordinates $x^1,...,x^n,y^1,...,y^m$ in 
${\mathbb{R}}^{n+m}$.} on ${\mathbb{R}}^n$:
\begin{equation}\label{10.15}
T(x,y)=\sum_{p\geq 
0}\frac{1}{p!}T^{i_1...i_p}(x)\frac{\partial}{\partial y^{i_1}}...
\frac{\partial}{\partial y^{i_p}}\delta(y)
\end{equation}
where $\delta$ is an $m$-dimensional Dirac function
\begin{equation}
\delta(y):=\delta(y^1)...\delta(y^m).
\end{equation}
Since the derivatives $\frac{\partial}{\partial y^j}$ mutually 
commute, $T^{i_1...i_p}$ is symmetrical in its indices, and the 
summation in (\ref{10.15}) is finite (from $p=0$ to $N$, for some 
$N\geq 0$.)
\par By integration one obtains
\begin{equation}
\int_{{\mathbb{R}}^{n+m}}\dd ^nx\,\,\,\dd ^my\,F(x,y)T(x,y)=\int_{{\mathbb{R}}^n}\dd ^nx\,LF(x,y)\mid_{y=0}
\end{equation}
with the differential operator
\begin{equation}
L=\sum_{p\geq 0} \frac{1}{p!}T^{i_1...i_p}(x)\frac{\partial}{\partial y^{i_1}}...
\frac{\partial}{\partial y^{i_p}}
\end{equation}
The analogy with the Gra\ss mann case (\ref{10.14}) is obvious.
\par In physical terms, (\ref{10.15}) means that $T$ is a sum of 
multiple sheets along ${\mathbb{R}}^n$, hence is localized in an 
infinitesimal neighborhood of ${\mathbb{R}}^n$ in 
${\mathbb{R}}^{n+m}$. By analogy we can assert that {\it the whole 
superspace ${\mathbb{R}}^{n|\nu}$ is an infinitesimal neighborhood 
of its body ${\mathbb{R}}^n$}.

\subsection{Exterior forms on a graded manifold}
We consider now forms and densities on a graded manifold $\dM$. 
Tensor calculus can be developed on a graded manifold in a more or 
less obvious way, taking into account the sign rules. For 
instance, corresponding to a local chart with coordinates 
$(x^A)=(x^a,\xi^\alpha)$, we introduce the differentials $\dd 
x^A$ and the vector fields $\partial_A$. The Lie derivative 
associated to $\partial_A$ acts on a superfunction $F(x,\xi)$ as the partial 
derivative $\frac{\partial F}{\partial x^A}$. By comparing the 
parities of $F$ and $\frac{\partial F}{\partial x^A}$, we conclude 
that the operator $\frac{\partial}{\partial x^A}$ increases the 
parity of $F$ by that of $x^A$, hence $\partial_A$ {\it has the 
same parity as $x^A$}. On the other hand, in the case of exterior 
differential forms, we require
\begin{equation}
\dd x^a\wedge\dd x^b=-\dd x^b\wedge\dd x^a
\end{equation}
for ordinary variables $x^a$ and a consistent extension is
\begin{equation}\label{10.4.2}
\dd x^A\wedge\dd x^B=(-1)^{(\tilde{A}+1)(\tilde{B}+1)}\dd x^B\wedge\dd 
x^A.
\end{equation}
Therefore {\it $x^A$ and $\dd x^A$ have opposite parities.}
\par We conclude
\begin{eqnarray}
\widetilde{\frac{\partial}{\partial x^a}}=0 & & \widetilde{\frac{\partial}{\partial 
\xi^\alpha}}=1\\
\widetilde{\dd x^a}=1 & & \widetilde{\dd \xi^\alpha}=0.
\end{eqnarray}
Starting from (\ref{10.4.2}), we generate the $p$-forms by products 
of $p$ forms of the type $\dd x^A$, that is 
\begin{equation}\label{10.4.4}
\omega=\frac{1}{p!}\omega_{A_1...A_p}\dd x^{A_1}\wedge...\wedge\dd 
x^{A_p}
\end{equation}
It can also be written as 
\begin{equation}
\omega=\sum_{q+r=p}\frac{1}{q!r!}\omega_{a_1...a_q\alpha_1...\alpha_r}\dd 
x^{a_1}\wedge...\wedge\dd x^{a_q}\wedge\dd 
\xi^{\alpha_1}\wedge...\wedge\dd \xi^{\alpha_r}
\end{equation}
with components antisymmetrical in the bosonic indices 
$a_1,...,a_q$ and symmetrical in the fermionic indices 
$\alpha_1,...,\alpha_r$. 
{\it The parity of $\omega$ is that of $q$, that is count the 
number of bosonic differentials $\dd x^a$.}
\par To change from the coordinate system $(x^A)$ to another one 
$(\bar{x}^A)$ introduce the Jacobian matrix by
\begin{equation}
\dd x^A=\dd \bar{x}^P\cdot {}_PJ^A
\end{equation}
where ${}_PJ^A=\partial x^A/\partial \bar{x}^P$ (this gives the 
correct sign for the partial derivative w.r.t. a fermionic 
variable $\xi^\alpha$). We calculate for instance
\begin{equation}
\dd x^A\wedge\dd x^B=(-1)^{\tilde{Q}(\tilde{A}+\tilde{P})}\dd 
\bar{x}^P\wedge\dd \bar{x}^Q\,{}_PJ^A\,{}_QJ^B
\end{equation}
The (total) differential $\dd F$ of a superfunction $F$ is defined 
invariantly by
\begin{equation}
\dd F=\dd x^A\cdot\frac{\partial F}{\partial x^A}
\end{equation}
For the parity, we get $\widetilde{\dd F}=\tilde{F}+1$, and this 
rule assigns parity 1 to the operator $\dd$. We then extend the 
differential $\dd$ to an exterior differential of forms by
\begin{equation}
\dd \omega=\frac{1}{p!}\dd \omega_{A_1...A_p}\wedge\dd 
x^{A_1}\wedge...\wedge\dd x^{A_p}
\end{equation}
for $\omega$ given by (\ref{10.4.4}). {\it The exterior 
differentiation is an operator of parity 1.}
\par Let $X$ be a supervector field with components ${}^AX$.
The contraction $\di_X\omega$ of a $p$-form $\omega$ is a $(p-1)$-form 
defined in components by
\begin{equation}
(\di_X\omega)_{A_2...A_p}=X^{A_1}{}_{A_1}\omega_{A_2...A_p}
\end{equation}
According to the rules of superalgebra the various components of $\omega$ 
and $X$ are related by 
\begin{equation}
X^A=(-1)^{\tilde{A}\tilde{X}}\,{}^AX\mbox{,~~~~~ }{}_{A_1}\omega_{A_2...A_p}=(-1)^{\tilde{\omega}\cdot\tilde{A_1}}\omega_{A_1A_2...A_p}.
\end{equation}
With this definition the parity is given by
\begin{equation}
\widetilde{\di_X\omega}=\tilde{\omega}+\tilde{X}+1
\end{equation}
and this assigns $\widetilde{\di_X}=\tilde{X}+1$ as the parity of 
the operator $\di_X$, hence the symbol $\di$ has to be considered 
as {\it odd}.
\par Finally, the Lie derivative acting on forms is the graded 
commutator
\begin{equation}
{\mathcal{L}}_X=[\di_X,\dd]
\end{equation}
a graded operator of the same parity as $X$.
We denote as before by ${\mathcal{A}}(\dM)$ the vectorspace of 
$p$-forms on $\dM$. We collect the definition of the various 
operators acting on forms:
\begin{eqnarray}
M(F):&{\mathcal{A}}^p(\dM)\rightarrow{\mathcal{A}}^p(\dM)&\mbox{  
multiplication by $F$}\\
\de(F):&{\mathcal{A}}^p(\dM)\rightarrow{\mathcal{A}}^{p+1}(\dM)&\mbox{  
multiplication by $\dd F$}\\
\di_X:&{\mathcal{A}}^p(\dM)\rightarrow{\mathcal{A}}^{p-1}(\dM)&\mbox{  
contraction by $X$}\\
\dd :&{\mathcal{A}}^p(\dM)\rightarrow{\mathcal{A}}^{p+1}(\dM)&\mbox{  
exterior derivative}\\
{\mathcal{L}}_X:&{\mathcal{A}}^p(\dM)\rightarrow{\mathcal{A}}^p(\dM)&\mbox{  
Lie derivative}
\end{eqnarray}
(where $F$ is a superfunction on $\dM$, and $X$ a supervectorfield). 
The graded commutators are 0 except the following ones:
\begin{eqnarray}
[\di_X,\de (F)] & = & M({\mathcal{L}}_XF)=[ {\mathcal{L}}_X,M(F) ] \\
{}[\di_X,\dd ] &=& {\mathcal{L}}_X \\
{}[\dd,M(f)]&=&\de(F) \\
{}[{\mathcal{L}}_X,{\mathcal{L}}_Y]&=&{\mathcal{L}}_{[X,Y]}\\
{}[{\mathcal{L}}_X,\de(F)]&=&\de({\mathcal{L}}_XF)\\
{}[{\mathcal{L}}_X,\di_Y]&=&\di_{[X,Y]}.
\end{eqnarray}
Since $\dd \dd=0$, we get an ascending complex of forms
\begin{equation}
{\mathcal{A}}^0(\dM)\stackrel{\dd}{\longrightarrow}{\mathcal{A}}^1(\dM)\stackrel{\dd}{\longrightarrow}{\mathcal{A}}^2(\dM)\stackrel{\dd}{\longrightarrow}...
\end{equation}
which is unbounded above if $\dM$ is not an ordinary manifold (that is $\nu >0$).

\subsection{Densities on a graded manifold}
We combine what we said above in the pure bosonic and the pure 
fermionic cases. We shall be sketchy and leave the technical 
details to a forthcoming publication.
\par A {\it scalar density} (of weight one) is given in coordinates 
by one component ${\mathfrak{F}}(x,\xi)$ but it can be viewed as an ordinary
tensor
\begin{equation}
{\mathfrak{F}}_{i_1...i_n}{}^{\alpha_1...\alpha_\nu}
\end{equation}
totally antisymmetric in $i_1,...,i_n$ and $\alpha_1,...,\alpha_\nu$ 
separately, with ${\mathfrak{F}}={\mathfrak{F}}_{1...n}{}^{1...\nu}$. In more intrinsic 
terms\,\footnote{~We use the abbreviation 
$\delta_\alpha=\partial/\partial \xi^\alpha$.}
\begin{equation}
{\mathcal{F}}=\dd x^1\wedge...\wedge\dd 
x^n\otimes\delta_1...\delta_\nu {\mathfrak{F}}(X,\xi)
\end{equation}
It behaves in such a tensorial way under a coordinate 
transformation where the fermionic variables transform linearly, 
but not under a general coordinate transformation. 
\par To integrate such a density, we take our advice from subsection 
\ref{10.1}:
\begin{equation}
\int_\dM{\mathcal{F}}=\int_{\dM_B}\omega
\end{equation}
where $\dM_B$ is the body of $\dM$ and the $n$-form $\omega$ is given 
by
\begin{equation}
\omega=\dd x^1\wedge...\wedge\dd 
x^n\left(\delta_1...\delta_\nu{\mathfrak{F}}(x,\xi)\right)
\end{equation}
this last expression being independent of $\xi$. Since we can 
multiply a density ${\mathcal{F}}$ by a superfunction $F$, we get 
an integration process
\begin{equation}
F\mapsto\int_\dM {\mathcal{F}}\cdot F
\end{equation}
on $\dM$. It is really an integro-differential process, and it can 
be split as follows:
\begin{enumerate}
\item a differential operator $P$ mapping superfunctions $F$ on 
$\dM$ to top-forms $P(F)$ on the body $\dM_B$;
\item integrating $P(F)$ on $\dM_B$.
\end{enumerate}
This splitting has an invariant meaning for manifolds split in 
their body and soul, and provides an alternative definition for 
the so-called Berezinian. 
\par  From the scalar densities, we construct a descending complex of 
densities
\begin{equation}
{\mathcal{D}}_0\stackrel{\nabla\cdot}{\longleftarrow}{\mathcal{D}}_1\stackrel{\nabla\cdot}{\longleftarrow}...
\end{equation}
with its cohort of operators $\de(F)$, $\di_X$, $M(F)$, 
${\mathcal{L}}_X$.

\part{Applications}
\pagestyle{myheadings}  
\markboth{\quad\small \em P. Cartier, C. DeWitt-Morette, M. Ihl and C. S{\"a}mann \hfill}{\hfill \small \em Supermanifolds -- application to supersymmetry\quad}

\section{$\Pi T\dM$ and $\Pi T^*\dM$}
\setcounter{equation}{0}

A special case of supermanifolds is obtained by changing the 
parity of the fiber coordinates of a vector bundle. We introduce 
the parity operator $\Pi$:
\vspace{0.2cm}
\begin{definition}
The parity operator $\Pi$ acts on a fiber bundle by changing the 
parity of the fiber coordinates.
\end{definition}
\vspace{0.2cm}
Given the tangent bundle $T\dM$ over an $n$-dimensional manifold 
which is locally described by $2n$ real coordinates 
$x^1,...,x^n,\dot{x}^1,...,\dot{x}^n$, $\Pi T\dM$ is a graded manifold of 
dimensions $(n,n)$ and has coordinates 
$x^1,...,x^n,\xi^1,...,\xi^n$ where $\xi^i$ are Gra\ss mann 
variables. Similarly, the cotangent bundle $T^*\dM$ has local coordinates 
$x^1,...,x^n,p_1,...,p_n$ so that $\Pi T^*\dM$ is locally described 
by $x^1,...,x^n,\pi_1,...,\pi_n$ where again the $\pi_i$ are 
Gra\ss mann variables.
\par The graded manifolds $\Pi T\dM$ and $\Pi T^*\dM$ have equally many 
even and odd dimensions, which is required for a linear description of 
supersymmetric systems.

\section{Supersymmetric Fock space}
\setcounter{equation}{0}

\subsection{Definition of a Fock space}
A Fock space is a Hilbert space ${\mathcal{H}}$ with a realization of the algebra:
\begin{eqnarray}
\hat{a}_i|0\rangle&:=&0\\
\left[\hat{a}_i,\hat{a}_j^\dag\right]_\mp=\hat{a}_i\hat{a}_j^\dag\mp\hat{a}_j^\dag\hat{a}_i&:=&\delta_{ij}.
\end{eqnarray}
The upper sign defines a bosonic Fock space, the lower sign a 
fermionic Fock space. We set $\hat{b}_i=\hat{a}_i$ in the bosonic 
and $\hat{f}_i=\hat{a}_i$ in the fermionic case. 
\par If both algebras together with the additional rules
\begin{eqnarray}
\left[\hat{b}_i,\hat{f}_j\right]&=&\left[\hat{b}_i,\hat{f}_j\right]_-:=0\nonumber\\
\left[\hat{b}_i,\hat{f}_j^\dag\right]&=&\left[\hat{b}_i,\hat{f}_j^\dag\right]_-:=0\nonumber
\end{eqnarray}
are simultaneously realized on a Hilbert space ${\mathcal{H}}$, we call ${\mathcal{H}}$ 
a supersymmetric Fock space.

\subsection{Holomorphic representation}
In the following section, $z$ and $\zeta$ denote ordinary complex 
and Gra\ss mann variables respectively\par 
A representation of the bosonic algebra\,\footnote{~We allow $N$ 
different degrees of freedom, i.e.~in physical terms e.g.~$N$ different momenta or 
positions on a lattice.} on ${\mathbb{C}}[z^1,...,z^N]$
is found by using the following definitions:
\begin{equation}
\hat{b}_i:=\frac{\partial}{\partial z^i}\mbox{ and 
}\hat{b}^\dag_i:=z^i\cdot
\end{equation}
The vacuum state $|0\rangle$ is represented by the function 
$f_0(z^1,...,z^N)=1$.
\par The Hilbert space ${\mathbb{C}}[z^1,...,z^N]$ is 
self-dual so that scalar product and dual product are identical:
\begin{equation}
\langle f|g\rangle=(f|g):=\int 
\dd z^1\dd \bar{z}^1...\dd z^N\dd \bar{z}^N 
{\mathrm{exp}}^{-z^1\bar{z}^1...-z^N\bar{z}^N}\overline{f(z)}g(z)
\end{equation}
The analogous representation of the fermionic algebra is found on 
the space of polynomials of $N$ complex Gra\ss mann variables ${\mathbb{C}}[\zeta^1,...,\zeta^N]$:
\begin{equation}
\hat{f}_i:=\frac{\partial}{\partial \zeta^i}\mbox{ and 
}\hat{f}^\dag_i:=\zeta^i\cdot
\end{equation} 
The vacuum state $|0\rangle$ is again the function 
$f_0(\zeta^1,...,\zeta^N)=1$.
The scalar product in ${\mathbb{C}}[\zeta^1,...,\zeta^N]$ is analogously:
\begin{equation}
(f|g):=\int 
\dd \zeta^1\dd \bar{\zeta}^1...\dd \zeta^N\dd \bar{\zeta}^N 
{\mathrm{exp}}^{-z^1\bar{\zeta}^1...-z^N\bar{\zeta}^N}\overline{f(\zeta)}g(\zeta)
\end{equation}
Using a superfunction on a space described by $N$ real and $N$ 
Gra\ss mann variables, we find a representation of the 
supersymmetric Fock space. Particularly, we can use the space of 
functions on $\Pi T\dM$: ${\mathbb{C}}[z^1,...,z^N,\zeta^1,...,\zeta^n]$.

\subsection{Representation by Forms and Densities}
There is an obvious one-to-one correspondence from ${\mathbb{C}}[z^1,...,z^N,\zeta^1,...,\zeta^n]$ to 
the space of (complexified) forms on a graded manifold $\Omega^\bullet(\Pi 
T\dM)$: One can substitute powers of $z^i$ by powers\,\footnote{~the 
product of forms being the wedge product} of $\dd \xi^i$ 
and powers of $\zeta^i$ by $\dd x^i$. We thus replace 
commuting and odd variables by one-forms of the same 
parity.
\par The operators in this representation become:
\begin{eqnarray}
\hat{b}_i:=\di (\partial_{\xi^i})&\mbox{ and 
}&\hat{b}^\dag_i:=\de (\xi^i)\\
\hat{f}_i:=\di (\partial_{x^i})&\mbox{ and 
}&\hat{f}^\dag_i:=\de (x^i)
\end{eqnarray}
It is easily verified, that the algebra is correct. Particularly.
\begin{equation}
[\di (\partial_{\xi^i}),\de (\xi^j)]_-=[\di (\partial_{x^i}),\de (x^j)]_+=\delta_{ij}
\end{equation}
This representation is obviously not self-dual. The dual 
representation is found on the space of densities on $\Pi T\dM$:
${\mathcal{D}}_\bullet(\Pi T\dM)$. If we write tensor densities 
formally as
\begin{equation}
{\mathfrak{F}}=\sqrt{{\mathrm{det}}\,g}\frac{\partial}{\partial 
x^{A_1}}\wedge...\wedge\frac{\partial}{\partial 
x^{A_m}}{\mathfrak{F}}^{A_1...A_m}
\end{equation}
then the representation by densities is obtained from the 
holomorphic representation by substituting powers of $z^i$ by 
powers of $(\partial/\partial \xi^i)$ and powers of $\zeta^i$ by 
powers of $(\partial/\partial x^i)$ with, what is somewhat 
unusual, the wedge product between the partial derivatives. 
\par The operators are again 
\begin{eqnarray}
\hat{b}_i:=\di (\partial_{\xi^i})&\mbox{ and 
}&\hat{b}^\dag_i:=\de (\xi^i)\\
\hat{f}_i:=\di (\partial_{x^i})&\mbox{ and 
}&\hat{f}^\dag_i:=\de (x^i)
\end{eqnarray}
which act in the coordinate expansion on 
${\mathfrak{F}}^{A_1...A_m}$ as described above.
\par The dual product is now obtained by contracting the tensor 
density with the form, which yields a scalar density, and 
integrating over the configuration space:
\begin{equation}
\langle \psi_1|\psi_2\rangle:=\int_\dM 
\langle{\mathfrak{F}}(\psi_1),\omega(\psi_2)\rangle\omega^{top}=\int_\dM \left({\mathfrak{F}}^{A_1...A_i}(\psi_1)
\omega_{\alpha_1...\alpha_j}(\psi_2)\right)\omega^{top}.
\end{equation}
This expression vanishes, unless the degree of the tensor 
density and the form are identical, i.e.~$i=j$.

\section{Dirac Matrices}
\setcounter{equation}{0}

Many authors have remarked the connection between the Dirac 
operators and the operators $\dd $ and $\delta$ acting on 
differential forms. Here are some supplementary remarks.
\par Consider a $D$-dimensional real vector space $V$ with a scalar 
product. Introducing a basis $e_1,...,e_D$ we represent a vector 
by its components $v=v^ae_a$ and the scalar product reads
\begin{equation}
g(v,w)=g_{ab}v^aw^b
\end{equation}
Let $C(V)$ be the corresponding Clifford algebra 
generated by $\gamma_1,...,\gamma_D$ subjected to the relations
\begin{equation}
\gamma_a\gamma_b+\gamma_b\gamma_a=2g_{ab}
\end{equation}
The dual generators are given by $\gamma^a=g^{ab}\gamma_b$ and 
\begin{equation}\label{15.3}
\gamma^a\gamma^b+\gamma^b\gamma^a=2g^{ab}
\end{equation}
where $g^{ab}g_{bc}=\delta^a{}_c$ as usual.
\par We define now a representation of the Clifford algebra $C(V)$ by 
operators acting on a Gra\ss mann algebra. Introduce Gra\ss mann 
variables $\xi^1,...,\xi^D$ and put
\begin{equation}
\gamma^a=\xi^a+g^{ab}\frac{\partial}{\partial \xi^b}
\end{equation}
Then the relations (\ref{15.3}) hold. In more intrinsic terms we 
consider the exterior algebra $\Lambda V^*$ built on the dual 
$V^*$ of $V$ with a basis $(\xi^1,...,\xi^D)$ dual to the basis 
$(e_1,...,e_n)$ of $V$. The scalar product $g$ defines an 
isomorphism $v\mapsto I_g v$ of $V$ with $V^*$ characterized by
\begin{equation}
\langle I_gv,w\rangle=g(v,w)
\end{equation}
Then we define the operator $\gamma(v)$ acting on $\Lambda V^*$ as 
follows
\begin{equation}
\gamma(v)\cdot\omega=I_gv\wedge\omega+\di (v)\omega
\end{equation}
where the contraction operator $\di (v)$ satisfies
\begin{equation}
\di (v)(\omega_1\wedge...\wedge\omega_p)=\sum_{j=1}^p(-1)^{j-1}
\langle\omega_j,v\rangle\omega_1\wedge...\wedge\hat{\omega}_j\wedge...\wedge\omega_p
\end{equation}
(The hat \,$\hat{}$\, means omitting the corresponding factor).
An easy calculation gives
\begin{equation}
\gamma(v)\gamma(w)+\gamma(w)\gamma(v)=2g(v,w)
\end{equation}
We recover $\gamma_a=\gamma(\de _a)$, hence 
$\gamma^a=g^{ab}\gamma_b$.
\par {\it The representation thus constructed is not the spinor 
representation} since it is of dimension $2^D$. Assume $D$ is even, 
$D=2n$, for simplicity. Hence $\Lambda V^*$ is of dimension 
$2^D=(2^n)^2$ and {\it the spinor representation should be a ``square 
root" of $\Lambda V^*$}.
\par Indeed, on $\Lambda V^*$ consider the operator $J$ given by
\begin{equation}
J(\omega_1\wedge...\wedge\omega_p)=\omega_p\wedge...\wedge\omega_1=(-1)^{p(p-1)/2}\omega_1\wedge...\wedge\omega_p
\end{equation}
and introduce the operators
\begin{equation}
\gamma^0(v)=J\gamma(v)J
\end{equation}
Since $J^2=1$, they satisfy the Clifford relations
\begin{equation}
\gamma^0(v)\gamma^0(w)+\gamma^0(w)\gamma^0(v)=2g(v,w)
\end{equation}
In components $\gamma^0(v)=v^a\gamma^0_a$ where 
$\gamma^0_a=J\gamma_aJ$. The interesting point is the commutation 
property\,\footnote{~This construction is reminiscent of Connes' 
description of the standard model in A.Connes, {\it G\'{e}om\'{e}trie noncommutative}, ch. 5, InterEditions, Paris, 1990.}
\begin{center}
{\it $\gamma(v)$ and $\gamma^0(w)$ commute for all $v$, $w$}
\end{center}
According to the standard wisdom of quantum theory, the degrees of 
freedom associated with the $\gamma_a$ decouple with the ones for 
the $\gamma^0_a$. Assume that the scalar are complex numbers, 
hence the Clifford algebra is isomorphic to the algebra of 
matrices of type $2^n\times 2^n$. Then $\Lambda V^*$ can be 
decomposed as a tensor square
\begin{equation}
\Lambda V^*=S\otimes S
\end{equation}
with the $\gamma(v)$ acting on the first factor only, and the 
$\gamma^0(v)$ acting on the second factor in the same way:
\begin{eqnarray}
\gamma(v)(\psi\otimes\psi')&=&\Gamma(v)\psi\otimes\psi'\label{15.12}\\
\gamma^0(v)(\psi\otimes\psi')&=&v\otimes\Gamma(v)\psi'
\end{eqnarray}
The operator $J$ is then the exchange
\begin{equation}\label{15.14}
J(\psi\otimes\psi')=\psi'\otimes\psi
\end{equation}
The decomposition $S\otimes S=\Lambda V^*$ corresponds to the 
formula
\begin{equation}
c_{i_1...i_p}=\bar{\psi}\gamma_{[i_1}...\gamma_{i_p]}\psi\hspace{0.5cm}(0\leq 
p\leq D)
\end{equation}
for the currents\,\footnote{~For $n=4$, this gives a scalar, a vector, a bivector, a pseudo-vector and a 
pseudo-scalar.} $c_{i_1...i_p}$ (by $[...]$ we denote 
antisymmetrization).
\par In differential geometric terms, let $(\dM^D,g)$ be a 
(pseudo-)Riemannian manifold. The Gra\ss mann algebra $\Lambda 
V^*$ is replaced by the graded algebra ${\mathcal{A}}(\dM)$ of 
differential forms. The Clifford operators are given by
\begin{equation}
\gamma(f)\omega=\dd f\wedge\omega+\di (\nabla 
f)\omega
\end{equation}
($\nabla f$ is the gradient of $f$ w.r.t. the metric $g$, 
a vector field). In components
$\gamma(f)=\partial_\mu f\cdot\gamma^\mu$ with 
\begin{equation}
\gamma^\mu=\de (x^\mu)+g^{\mu\nu}\di \left(\frac{\partial}{\partial 
x^\nu}\right)
\end{equation}
The operator $J$ satisfies
\begin{equation}
J(\omega)=(-1)^{p(p-1)/2}\omega
\end{equation}
for a $p$-form $\omega$. To give a spinor structure on the 
riemannian manifold $(\dM^D,g)$ (in the case $D$ even) is to 
give a splitting\,\footnote{~$T^*_{\mathbb{C}}\dM^D$ is the 
complexification of the cotangent bundle. We perform this complexification to avoid irrelevant discussions on the 
signature of the metric.}
\begin{equation}
\Lambda T^*_{\mathbb{C}}\dM^D\simeq S\otimes S
\end{equation}
satisfying the analogous of relations (\ref{15.12}) and 
(\ref{15.14}). The Dirac operator $\slasha{\partial}$ is then characterized 
by the fact that $\slasha{\partial}\times 1$ acting on bispinor fields 
(sections of $S\otimes S$ on $\dM^D$) corresponds to 
$\dd +\delta$ acting on (complex) differential forms, that 
is on (complex) superfunctions on $\Pi T\dM^D$.

\section*{Acknowledgments}
\addcontentsline{toc}{section}{\numberline{}Acknowledgments}
The final version of this paper has been written at the {\it 
Institut des Hautes Etudes Scientifiques}. One of us (C.S.) was 
given the opportunity of staying at the IHES between his two 
recent assignments (at the University of Texas at Austin, as a 
W\"urzburg exchange student, and at the Ecole Normale Sup\'{e}rieure 
(Paris) as a DMA visitor). The hospitality of the IHES is deeply 
appreciated.

\section*{Note added in proof}
\addcontentsline{toc}{section}{\numberline{}Note added in proof}
After finishing this manuscript, we looked for a couple of references to
the work of D.~Leites, aware of his work, but not familiar with it. We
discovered 103 references, and the monumental "Seminar on
Supermanifolds" (SOS) over 2000 pages written by D.~Leites, colleagues,
students and collaborators from 1977 to 2000. It includes in particular
a contribution by V. Shander ``Integration theory on 
supermanifolds''
Chapter 5, pp. 45-131. Needless to express our regret for discovering
this gold mine only now. For those who also have missed it, we give one
access to this large body of information: {\it mleites@matematik.su.se}.
The first definition of supervarieties (due to Leites) appeared in 1974 in Russian ``Spectra of
graded commutative rings'', Uspehi Matem. Nauk., 30 n°3, 209-210.
Early references to supersymmetry can be found in Julius Wess and Jan
Bagger "Supersymmetry and Supergravity" Princeton University Press 1973
and {\it The Many Faces of the Superworld} Yuri Golfand Memorial Volume,
ed. by M. Shifman, World Scientific, Singapore, 1999. See also Deligne P. et al (eds.) 
{\it Quantum fields and strings: a course for mathematicians}. 
Vol. 1, 2. Material from the Special Year on 
Quantum Field Theory held at the Institute for Advanced Study, 
Princeton, NJ, 1996--1997. AMS, Providence, RI; Institute for 
Advanced Study (IAS), Princeton, NJ, 1999. Vol. 1: xxii+723 pp.; Vol. 2: pp. i--xxiv and 
727--1501.

\appendix
\section{Appendix: Complex Conjugation of Gra\ss mann Quantities}
\setcounter{equation}{0}
 
\begin{center}
Maria E. Bell\\
\small{\it
University of Texas,\\ Department of Physics and Center for Relativity\\
Austin, TX 78712, USA}
\end{center}
\subsection{Supernumbers}
B.S. DeWitt considers the basic Gra\ss mann variables $\xi^i$ used 
to generate supernumbers as real. Nevertheless, by allowing in the 
expansion of a supernumber, namely
\begin{equation}\label{A1}
\psi=c_0+c_i\xi^i+\frac{1}{2!}c_{ij}\xi^i\xi^j+...
\end{equation}
the coefficients $c_0$, $c_i$, $c_{ij}(=-c_{ji})$ to be complex 
numbers, we define {\it complex} supernumbers. By separating in 
each coefficient real and imaginary part, we can write
\begin{equation}\label{A2}
\psi=\rho+\di \sigma
\end{equation}
where both $\rho$ and $\sigma$ have real coefficients. {\it In our 
conventions, a supernumber $\psi$ is real iff all its coefficients 
$c_{i_1...i_p}$ are real numbers.} In the decomposition (\ref{A2}) 
$\rho$ is the real part of $\psi$ and $\sigma$ its imaginary part. 
We {\it define} complex conjugation by
\begin{equation}
(\rho+\di \sigma)^*=\rho-\di \sigma
\end{equation}
for $\rho$, $\sigma$ real.
\par According to these rules, the generators $\xi^i$ are real, and 
sum and product of real supernumbers are real. Furthermore
\begin{eqnarray}
(\psi+\psi')^*&=&\psi^*+\psi^{\prime *} \label{A4}\\
(\psi\psi')^*=\psi^*\psi^{\prime *}&=&(-1)^{\tilde{\psi}\tilde{\psi}'}\psi^{\prime *}\psi^*\label{A5}\\
\psi\mbox{ is real iff }\psi^*&=&\psi\label{A6}
\end{eqnarray}
According to B.S. DeWitt's conventions, the rules (\ref{A4}) and 
(\ref{A6}) still hold, but (\ref{A5}) is replaced by
\begin{equation}
(\psi\psi')^*=(-1)^{\tilde{\psi}\tilde{\psi}'}\psi^*\psi^{\prime *}=\psi^{\prime *}\psi^*\label{A5BS}
\end{equation}
As a consequence, the product of two real supernumbers is purely 
imaginary.
\subsection{The Supertranspose}
We denote by $A^T$ the transpose of a matrix $A$. If $A$ and $B$ 
are matrices with supernumbers as entries, all of parity $a$ ($b$) 
for $A$ ($B$), then the product rule reads as
\begin{equation}\label{A7}
(AB)^T=(-1)^{ab}B^TA^T.
\end{equation}
We define now the {\it supertranspose} $K^{sT}$ of a graded matrix $K$ 
(our conventions agree with those of B.S. DeWitt). In terms of 
components, we have
\begin{equation}\label{A8}
{}_i(K^{sT})^j:=(-1)^{(\tilde{K}+\tilde{i})(\tilde{j}+\tilde{i})}\:{}^jK_i.
\end{equation}
In block form, we get
\begin{equation}
K=\left(
\begin{array}{*{2}{c}}
A & C \\
D & B \\
\end{array}
\right)
\end{equation}
where all elements of $A$ and $B$ are of parity $\tilde{K}$, while 
those of $C$ and $D$ are of parity $\tilde{K}+1$. Then 
\begin{equation}\label{A10}
K=\left(
\begin{array}{*{2}{c}}
A^T & (-1)^{\tilde{K}+1}D^T \\
(-1)^{\tilde{K}}C^T & B^T \\
\end{array}
\right)
\end{equation}
From the definition (\ref{A8}) (or from the definition (\ref{A10}) 
using the rule (\ref{A7}), one derives
\begin{equation}\label{A11}
(KL)^{sT}=(-1)^{\tilde{K}\tilde{L}}L^{sT}K^{sT}.
\end{equation}
\subsection{The Superhermitian Conjugate}
The {\it superhermitian conjugate} of a graded matrix $K$ is 
defined by
\begin{equation}
K^{sH}:=(K^{sT})^*=(K^*)^{sT}
\end{equation}
In this formula, we use the complex conjugate matrix $K^*$, 
obtained by taking the complex conjugate of every entry of 
$K$. From the rule (\ref{A5}), one derives immediately
\begin{equation}\label{A13}
(KL)^*=K^*L^*
\end{equation}
for graded matrices $K$ and $L$. Combining the rules (\ref{A11}) 
and (\ref{A13}), we immediately get
\begin{equation}\label{A14}
(KL)^{sH}=(-1)^{\tilde{K}\tilde{L}}L^{sH}.K^{sH}
\end{equation}
in complete agreement with Koszul's parity rule.
\par Conversely, if formula (\ref{A14}) is universally valid, it 
applies to $1\times1$ matrices, that is to supernumbers; hence we 
are back to (\ref{A5}).
\subsection{Graded operators}
The rule (\ref{A5}) together with its implication for hermitian 
conjugation applies to Gra\ss mann operators on a Hilbert space.
\\[0.2cm]
{\it Example}: Graded operators on Hilbert spaces.
\\[1mm]Let $|\Omega\rangle$ be a simultaneous eigenstate of 
$Z$ and $Z'$ with eigenvalues 
$z$ and $z'$.
\begin{equation}\label{A15}
ZZ'|\Omega\rangle=Zz'|\Omega\rangle=(-1)^{\tilde{Z}\tilde{z}'}z'Z|\Omega\rangle=(-1)^{\tilde{Z}\tilde{z}'}z'z|\Omega\rangle=zz'|\Omega\rangle
\end{equation}
since, it is clear from the eigenvalue equation 
$Z|\Omega\rangle=z|\Omega\rangle$ that an 
operator and its eigenvalue have the same parity.
\par The hermitian conjugate of the eigenvalue equation (\ref{A15}) is
\begin{equation}
(-1)^{\tilde{Z}'\widetilde{|\Omega\rangle}+\tilde{Z}\widetilde{|\Omega\rangle}+\tilde{Z}\tilde{Z}'}\langle\Omega|Z^{\prime sH}Z^{sH}=
(-1)^{\tilde{z}'\widetilde{|\Omega\rangle}+\tilde{z}\widetilde{|\Omega\rangle}+\tilde{z}\tilde{z}'}\langle\Omega|z^{\prime *} z^*
\end{equation}
On the other hand, using the argument leading to (\ref{A15})
\begin{eqnarray}
\langle\Omega|Z^{\prime sH}Z^{sH}&=&\langle\Omega|z^{\prime *}Z^{sH}=\langle\Omega|Z^{sH}z^{\prime *}(-1)^{\tilde{z}^{\prime *}\tilde{Z}}\nonumber\\
&=&\langle\Omega|z^*z^{\prime *}(-1)^{\tilde{z}^{\prime *}\tilde{Z}}=\langle\Omega|z^{\prime *}z^*
\end{eqnarray}

\section*{References}
\addcontentsline{toc}{section}{\numberline{}References}

\end{document}